\documentclass[useAMS,usenatbib]{mn2e}

\usepackage{lscape}
\usepackage{graphicx}
\usepackage{subfigure}

\newcommand\logmdot {$\log \dot{M}$}
\usepackage{multirow}

\title[Brown dwarf discs]{On the properties of discs around accreting brown dwarfs}  \author[Nathan J.  Mayne and Tim J. Harries]{Nathan J. Mayne\thanks{E-mail: nathan@astro.ex.ac.uk
    (NJM)} and Tim J. Harries\\
  School of Physics,
  University of Exeter, Stocker Road, Exeter, EX4 4QL.\\
}
 
\begin{document}

\date{Accepted ?. Received ?; in
  original form ?}

\pagerange{\pageref{firstpage}--\pageref{lastpage}} \pubyear{2010}

\maketitle

\label{firstpage}

\begin{abstract}

  We present a grid of models of accreting brown dwarf systems with
  circumstellar discs. The calculations involve a self-consistent
  solution of both vertical hydrostatic and radiative equilibrium
  along with a sophisticated treatment of dust sublimation. We have
  simulated observations of the spectral energy distributions and
  several broadband photometric systems. Analysis of the disc
  structures and simulated observations reveal a natural dichotomy in
  accretion rates, with \logmdot $>-$9 and $\leq -$9 classed as
  extreme and typical accretors respectively.  Derivation of ages and
  masses from our simulated photometry using isochrones is
  demonstrated to be unreliable even for typical accretors. Although
  current brown dwarf disc candidate selection criteria have been
  shown to be largely reliable when applied to our model grid we
  suggest improved selection criteria in several colour indices. We
  show that as accretion rates increase brown dwarf disc systems are
  less likely to be correctly identified. This suggests that, within
  our grid, systems with higher accretion rates would be
  preferentially lost during brown dwarf target selection. We suggest
  that observations used to assert a $\dot{M}\propto M_*^2$
  relationship may contain an intrinsic selection bias.

\end{abstract}

\begin{keywords}
  stars:evolution -- stars:formation -- stars: pre-main-sequence --
  techniques: photometric -- catalogues -- (stars) Hertzsprung-Russell
  H-R diagram
\end{keywords}

\section{Introduction}
\label{intro}

There is now strong evidence that as brown dwarfs (BDs) form they pass
through a classical T~Tauri star (CTTS) phase, during which they
possess a flared, dusty circumstellar disc from which they are
actively accreting material
\citep{jayawardhana_2003,mohanty_2004}. The accretion is thought to
proceed via a magnetically-controlled funnel flow mechanism in which
material from a truncated inner disc boundary falls onto the surface
of the star along magnetic field lines
\citep{camenzind_1990,koenigl_1991,muzerolle_2003,mohanty_2008}.

The interpretation of the spectral energy distributions (SEDs) of
pre-main-sequence (pre-MS) stars involves trying to distinguish the
various contributions to the continuum from the hot spots at the base
of the accretion flow, the photospheric flux, and the near-IR flux
from the dusty inner disc. The complexity of the interplay between
these contributions is exacerbated by other geometrical effects such
as the inclination (which changes the projected area of the inner disc
wall visible to the observer) and the outer disc structure
\citep[which may obscure the inner disc for high
inclinations][]{walker_2004,tannirkulam_2007}.

In surveys, particularly those attempting to discern the pre-MS disc
fraction disentangling of the SED components usually takes place using
broad-band photometric measures and cuts in colour-colour or
colour-magnitude space
\citep{luhman_2005b,luhman_2008,gutermuth_2008,monin_2010}.  The
important quantity here is the separation in wavelength between the
emission peaks for the stellar and thermal disc components, as it is
these components which must be isolated. For BD disc (BDD hereafter)
systems it is likely that difficulties detecting the disc will be
exacerbated by the lower temperatures of the photosphere and therefore
the smaller separation in wavelength from the thermal disc emission
component, compared to CTTS systems. Therefore, to derive disc
fractions for BD populations we require detailed comparison models
with which to guide disc candidate selection.

Comparison of accretion rates across the pre-MS mass spectrum has
indicated that the accretion rate is strongly correlated with pre-MS
mass, \citep{muzerolle_2003,natta_2004,natta_2006}, with an
approximate form $\dot{M}\propto M^2_*$. Since in the
canonical picture the accretion rate is driven by the disc viscosity,
and should be independent of the mass of the central object, this
correlation is somewhat surprising. There is some danger that the
correlation is the result of, or at least strengthened by, the
presence of observational biases \citep{clarke_2006}. At the high-mass
(CTTS) end of the mass spectrum the lowest accretion-rates cannot be
measured via continuum methods since the excess is too small in
contrast with the photospheric emission. The emission lines will also
be weak, and indeed the H$\alpha$ equivalent width (EW) may be less
than the 10\AA\ that traditionally demarcates classical from
weak-lined T Tauri stars \citep[although this boundary may well be
spectral type dependent,][]{navascues_2003}. Such objects should show
doppler-broadened profiles, but the line wings will be weak, and the
presence of underlying H$\alpha$ absorption may become dominant. Thus
there may well be a population of low-accretion-rate CTTS which are
current missing from the surveys.

There are also risks of observational biases at the low-mass end of
the correlation. Radiative-transfer modelling of H$\alpha$ emission
from accreting BDs requires high temperatures ($>$\, 10\,kK) in the
accretion funnels in order to recover the level of emission that is
observed \citep{muzerolle_2003,natta_2004,natta_2006}. Although
cooling rate arguments can be invoked to explain the presence of such
high temperatures, the line ratios of Pa$\beta$ and Br$\gamma$
indicate that the hydrogen emission may not arise in the funnel flows
\citep{gatti_2006}. However, recently some continuum measurements of
accreting BDs have been conducted \citep{herczeg_2009}, and these
broadly support the H$\alpha$ rates.

Nonetheless it is worth considering the observed effect that mass
accretion at typical CTTS rates onto a BD might have. Parity between
the BD photospheric luminosity and the accretion luminosity occurs at
relatively low accretion rates \citep{clarke_2006} and such objects
may not be detected as BDs at all in photometric surveys. Even before
such an extreme case is reached, the additional luminosity provided by
the accretion should have measurable effect on the BD
colours. Furthermore the additional flux will be reprocessed by the
disc, altering its scaleheight and possibly the shape of its inner
edge.

Here we present a grid of models of BDD systems, including a
self-consistent treatment of the photospheric and accretion luminosity
sources and the interaction of that flux with the circumstellar disc.
We investigate the impact of the luminosity of the central source on
location and shape of the disc inner rim, as well as the large-scale
structure and flaring of the outer disc. We construct synthetic
colour-colour and colour-magnitude diagrams in order to examine the
efficacy of the photometric selections used to isolate brown dwarfs
and measure disc fractions
\citep{luhman_2005b,luhman_2008,gutermuth_2008,monin_2010}.

The complete model grid, with derived photometry and isochrones, is
available online through our browsing
tool\footnote{http://bd-server.astro.ex.ac.uk/} which is described in Appendix \ref{website}.

\section{Model}
\label{model}

In this section we detail the physical model adopted and assumptions
made (Section \ref{physics}), then explain key elements of the
radiative transfer code (Section \ref{torus}). Then we discuss the
derived values, such as broadband photometric magnitudes and colours
(in Section \ref{derived}). Some internal consistency checks are
given in Appendix \ref{consistency}.

\subsection{Physical model and assumptions}
\label{physics}

\subsubsection{Photospheric flux}
\label{phot_flux}

In order to model an accreting BDD system one must first model the
underlying photospheric flux. We have adopted a BD stellar interior
and atmospheric model grid and have then constructed the total
photospheric flux for any input value of stellar age and mass by
interpolating for surface gravity ($\log g$), effective temperature
($T_{\rm eff}$), radius ($R_*/R_{\odot}$) and luminosity
($L_*/L_{\odot}$). These values were then used to interpolate
atmospheric spectra for flux (ergs s$^{-1}$cm$^{-2} {\rm \AA}^{-1}$)
from 1200 to 2$\times$10$^7$ \AA. The spectra were subsequently
resampled onto 200 logarithmically spaced points. Careful inspection
ensured that no spectral features were removed during resampling. The
stellar interior models used for this study are the `DUSTY00' models
of \cite{chabrier_2000} combined with the `AMES-Dusty', atmospheric
models of \cite{chabrier_2000}, which are all available
online\footnote{http://perso.ens-lyon.fr/france.allard/}. For our
$T_{\rm eff}$ range of $\approx$\,3000\,K\,$<T_{\rm eff}<$1600\,K the
AMES-Dusty atmospheres are the most applicable (2700\,K$>T_{\rm
  eff}>$1700\,K). We did try including dynamic application of
atmospheres based on the derived $T_{\rm eff}$, i.e. using AMES-Cond
for $T_{\rm eff}<$1700\,K, but this resulted in large discontinuities
between the model atmospheres and resulting spectra. Additionally, for
the higher temperatures in our range, \cite{martin_2000} have found
the NextGen models to be more applicable for $\sim T_{\rm
  eff}>2300$K. However, as with the lower boundary this only applies
to the edge of our temperature range and we feel it is better to use a
consistent set of atmosphere and interior models across our
grid. Since this only affects stars at the very edge of our
temperature range, i.e. for the oldest and lowest mass objects (for
the AMES-Cond case), we have adopted the AMES-Dusty models throughout.

\subsubsection{Accretion flux}
\label{acc_flux}

We assumed blackbody emission for the accretion flux.  The selected
accretion rate was used to derive an accretion luminosity ($L_{\rm
  acc}$), where the material was modelled as free-falling from the
disc inner edge onto the surface of the star. $L_{\rm acc}$ is
calculated according to,
\begin{equation}
{L_{\rm acc}=\frac{GM_*\dot{M}}{R_*} \left( 1-\frac{R_*}{R_{\rm inner}}\right)},
\label{Lacc}
\end{equation}
where $M_*$ is the stellar mass, $\dot{M}$ the mass accretion rate,
$R_*$ the stellar radius and $R_{\rm inner}$ the radius of the disc inner
boundary.

The initial inner disc radius was set to be the co-rotation radius
(this is discussed in more detail in Section
\ref{disc_parameters}). During the radiative transfer simulations of
the disc the final inner dust-disc radius may be beyond the
co-rotation radius due to dust sublimation effects (see Sections
\ref{dust_edge} and \ref{inner_edge} for an explanation). Once the
accretion luminosity was derived, an adopted areal coverage ($A$),
over the stellar surface, was used to calculate an effective
temperature ($T_{\rm acc}$), for the accretion `hot' spot, where
\begin{equation}
{T_{\rm acc}=\left(\frac{L_{\rm acc}}{4\pi R^2_{*} \sigma A}\right)^{\frac{1}{4}}}.
\label{Tacc}
\end{equation}

Finally, a blackbody flux distribution is generated at $T_{\rm acc}$
and added onto the intrinsic stellar photospheric flux. In general,
one would expect this to be an overestimate of the accretion flux, as
for BDs large convective zones are expected on the stellar surface,
and some of the accretion energy may act to further drive these
convective currents, meaning flux is lost. It is worth noting however
that observationally UV excesses are often used to recreate and then
subtract an assumed accretion flux using a blackbody flux curve, which
is essentially the reverse of this method.

\subsubsection{Disc parameters}
\label{disc_parameters}

In this study we assume that accretion from the central star occurs
along magnetically channelled columns from the inner disc boundary.
For CTTS stars, \cite{bouvier_2007} show that the magnetic truncation
radius ($R_{\rm mag}$) is less than the co-rotation radius ($R_{\rm
  co}$), where the angular Keplerian velocity of the disc is equal to
the surface angular velocity of the central star.  Calculations of the
magnetic truncation radius depend on derivations of the surface
magnetic field \citep{koenigl_1991}. This is currently unavailable for
BDs, due to an increased number of molecular species obscuring
the Zeeman splitting signatures that are normally used to derive
stellar surface magnetic fields. Therefore, for our model grid we have
adopted an initial inner disc radius as the co-rotation radius,
\begin{equation}
{R_{\rm inner}=\left({GM_*\tau^2\over{4\pi ^2}}\right)^{1\over{3}}},
\label{inner_eq}
\end{equation}
where $\tau$ is the stellar rotation period and $R_{\rm inner}$ is the
inner radius. This is effectively adopting a disc-locking mechanism
(without associated angular momentum loss), as for disc-locked stars,
$R_{\rm mag}\approx R_{\rm co}$,
\citep{koenigl_1991,shu_1994}. Therefore, for our model simulations,
this inner edge radius is dependent on, and derived from, the adopted
value of the rotational period for the central star, as well as being
weakly dependent on the stellar mass. As discussed in Section
\ref{intro}, the inner disc can be cleared through a number of
mechanisms, including binarity or giant planet formation,
photoevaporation or photoionisation of the disc and dust grain growth
or settling. For BDD systems where a disc is modelled a treatment of
dust sublimation is included (discussed in Section \ref{dust_edge}).
However, the effects of binarity or giant planet formation are
neglected. Further to the stellar mass and period required prior to
calculation of the inner disc radius, we require a disc mass (in
stellar masses).

Although we solve for the vertical disc structure, the radial
structure of the disc is a free parameter. We assume that the surface
density varies as $\Sigma(r) \propto r^{-1}$.

For this work the disc outer edge was set at 300\,AU, this was chosen
as a maximum size of the circumstellar
disc. \cite{bouy_2008} have shown that the disc outer radius
has little effect on the resulting SED. However, in our subsequent
paper we will include models for outer radii of 100\,AU and plan to
extend this parameter range further to smaller values of the outer
radius in the future.

\subsubsection{Naked and disc systems}
\label{naked_BDD}

The combined (accretion plus photosphere) SED is then used as a
boundary condition for the {\sc torus} radiative transfer code and
as a benchmark set of SEDs to model `naked' BD systems. The set of
`naked' photospheres (plus accretion) are diluted by the factor
$(R_*/{\rm distance})^2$ to a distance of 10\,pc. The `negligibly' accreting,
`naked' stars can be used to produce absolute magnitude (and intrinsic
colour) isochrones for comparison. The remainder are used to
model systems showing active accretion where no disc is detected. For
instance, \cite{kennedy_2009} find 43 stars within their sample
are actively accreting whilst no disc is detected (out of a total
sample of 1253).

In summary the key required input variables to setup the model grid
are as follows: stellar age and mass (which are used to derive the
stellar flux) and accretion rate, areal coverage and rotation period
(which are used to determine the the accretion flux and temperature).
The disc parameters are set by a disc mass (expressed as a fraction of
the BD mass), and a power-law distribution for the disc surface
density.

\subsection{Radiative transfer code: {\sc torus}}
\label{torus}

We have used the {\sc torus} radiative transfer code which was
originally described by \cite{harries_2000} and was subsequently
updated by \cite{harries_2004} and \cite{kurosawa_2004}. {\sc torus}
uses the method of \cite{lucy_1999} to solve radiative equilibrium on
an AMR mesh. The code also self-consistently solves the equation of
vertical hydrostatic equilibrium and dust sublimation for the disc
\citep[described in][]{tannirkulam_2007}. The {\sc torus} code has
been extensively benchmarked \citep[for radiative transfer in discs
against][]{pinte_2009}.

\subsubsection{Dust Size distribution}
\label{dust_size}
 
\begin{figure}
  \includegraphics[scale=0.3,angle=90]{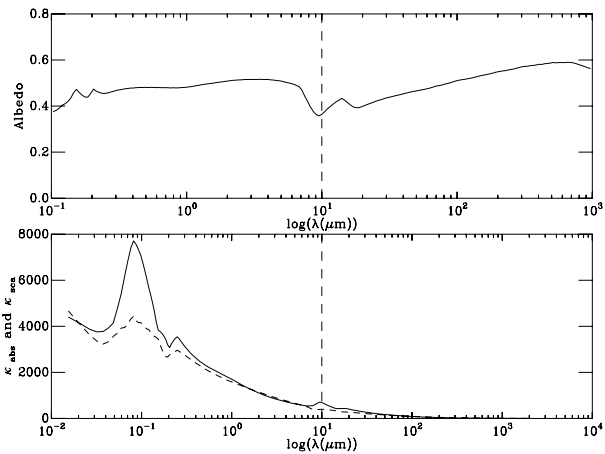}
 \caption{Figures of the albedo (top panel), and scattering
   (dashed line) and absorption (solid line) opacities (bottom
     panel) against log($\lambda$) (in $\mu$m) for our adopted dust
   population. For both panels the vertical dashed line is
   plotted at 10\,$\mu$m to highlight the silicate features.
   \label{albedo}}
\end{figure}

We have adopted a similar dust model to \cite{wood_2002} 
with the size distribution of dust particle given by
\begin{equation}
\label{particle_dist}
n(a)da=C_ia^{-q}\times exp^{[-(a/a_c)^p]}da
\end{equation}
where $n(a)da$ is the number of particles of size $a$ (within the
increment $da$), $a_{\rm c}$ is the characteristic particle size, with
$p$ and $q$ simply used to control the shape of the
distribution. $C_i$ controls the relative abundance of each
constituent species ($i$) in the dust. \cite{wood_2002}
found that simulated SEDs fit observed SEDs better using this adjusted
size distribution for the dust particles, as opposed to a simple power
law. The best fitting values found in \cite{wood_2002} were
$q=$3.0, $p=$0.6 with $a_{\rm c}=$50\,$\mu$m, also with an associated
maximum and minimum grain size of $a_{\rm min}=$5 nm and $a_{\rm
  max}=$1mm. A slightly steeper power law dependence, with $q=$3.5
(corresponding to the canonical MRN distribution) is more widely
adopted
\citep{bouy_2008,morrow_2008,pascucci_2008}.
Here we have adopted the values of \cite{wood_2002}
except for the parameter $q$ where we have used $q=$3.5.
\cite{wood_2002} calculated the $C_i$ values for amorphous
carbon and silicon by requiring the dust to deplete a solar abundance
of either component completely \citep[using abundances
  from][]{anders_1989,noels_1993}. We have set
$C_i=$1 and adjusted the species using a grain fractional abundance
\cite[an equivalent process to that of][]{wood_2002},
however we have adjusted these grain fractions using the updated solar
abundances of \cite{asplund_2006}. The resulting difference in
opacity between grain fractions matching the work of
\cite{wood_2002} and the new grain fractions is only a
slight enhancement of the silicate feature (due to the relative
abundance of silicon increasing) which has little effect on the
resulting SEDs. Figure \ref{albedo} shows the resulting albedo, and
scattering and absorption opacities, for our dust population, with a
the vertical dashed line showing 10\,$\mu$m.

\subsubsection{Dust sublimation and the inner disc edge}
\label{dust_edge}

As our models assume magnetic truncation at the co-rotation radius, we
have generated an inner hole. This also means that the disc will have
an inner wall at this radius. Evidence for inner walls in
circumstellar discs is apparent from the SEDs of disc systems, where a
peak in emission is found between 2 and 3\,$\mu$m. The temperatures
reached by such inner walls are expected to generate thermal flux
contributions within this wavelength range \citep{dullemond_2001}. The
circumstellar disc receives direct photospheric radiation and is
strongly heated. This leads to an increased scaleheight, or `puffing
up' of the inner disc rim, and subsequent shadowing of the region
immediately beyond it \citep{dullemond_2001}. The density-dependent
nature of the dust sublimation temperature \citep{pollack_1994} may
also play a role in shaping the inner rim
\citep{isella_2005,tannirkulam_2007}\footnote{Although we recognise
  that the inner hole in protostellar discs may be the result of giant
  planet formation or photoevaporation \citep{dahm_2009,najita_2007},
  these two mechanisms should result in a negligible mass-accretion
  rate \citep{dahm_2009} and therefore would represent a distinct
  evolutionary state of brown dwarf disc (BDD) systems to that of the
  CTTS-like phase we are considering here.}.

For low-luminosity systems the dust sublimation radius ($R_{\rm sub}$)
will be coincident with the co-rotation radius. However should the
combined photospheric and accretion luminosities be sufficient the
inner disc will be heated sufficiently that the dust close to the
co-rotation radius will be sublimated, and we must account for this in
our models. Furthermore, since the dust sublimation temperature has a
density dependence \citep{pollack_1994}, it is both the location and
{\em shape} of the inner wall that can change
\citep{isella_2005,tannirkulam_2007}.

Our treatment of dust sublimation is similar to that detailed in
\cite{tannirkulam_2007}, but with some enhancements to ensure a swift
convergence of the sublimated rim. The dust sublimation proceeds as
follows: An initial temperature distribution is found for the
optically-thin limit by setting the global dust-to-gas ratio to a tiny
value. Subsequent radiative equilibrium iterations are performed using
the adopted dust-to-gas ratio of 0.01, but limiting the maximum
optical depth across a given cell to $\tau_{\rm max}$. Cells whose
temperature exceeds the local dust sublimation temperature have their
dust-to-gas ratio set to zero. Radiative equilibrium iterations and
sublimation sweeps are performed at $\tau_{\max} = $0.1, 1, and 10,
with a final iteration of $\tau_{\max} = \infty$. Adequately solving
the radiative-equilibrium necessitates resolving the disc's effective
photosphere, and adaptive mesh refinement is used to split the grid at
the optically-thin/optically-thick boundary in order that the maximum
cell size at this boundary is $<$\,1 at a wavelength of
5500\,\AA. Once a self-consistent sublimated rim has been determined,
the equation of hydrostatic equilibrium is solved throughout the disc,
and the sublimation iterations are restarted (the change in the
density structure from the hydrostatical equilibrium step naturally
feeds back into the shape of the inner rim). Five hydrostatic
equilibrium steps are performed to ensure convergence, although a
stable disc structure is normally found after three such
iterations. For a more sophisticated, and chemically more complex,
study of dust sublimation albeit for higher mass objects see
\cite{kama_2009}.

For our model grid the gas population is assumed to be essentially
static with a zero optical depth.

\subsubsection{SEDs}
\label{seds}

Once the radiative transfer code was completed simulated SEDs were
generated. These SEDs can be generated for any distance and for any
system inclination. For our models we have set the distance, to 10\,pc
to create absolute flux SEDs and selected ten inclinations equi-spaced
in $\cos i$ (where $i$=0, 27, 39, 48, 56, 64, 71, 77, 84 and
90$^{\circ}$). A further useful feature of the {\sc torus} code is
that the emitted photon packets which contribute to the SED are tagged
on their way to the observer. These tags separate the packets into
four groups. Firstly, packets are separated by source into thermal
(disc) or stellar groups. These groups are then subdivided into those
which reach the observer either directly or after scattering. The
resulting SEDs are discussed in Section \ref{seds}.

\subsection{Photometric systems and derived quantities}
\label{derived}

Many observational studies use non-spectroscopic data to derive the
pertinent parameters. Therefore in order to examine the practical
effects in the `observational plane' we have used the SEDs to produce
broadband photometric magnitudes and subsequently colours. Broadband
magnitudes were also derived in a large range of other filter sets not
used explicitly in the analysis within this paper. These magnitudes
are available online\footnote{http://bd-server.astro.ex.ac.uk/} and
are briefly discussed in Appendix \ref{website}. In addition,
monochromatic fluxes have been derived for all filters, and again are
available online and discussed in Appendix \ref{website}.

In order to derive broadband photometric magnitudes and colours the
SEDs of either the disc or naked systems were folded through the
filter responses of the required photometric system. As the fluxes in
all cases are absolute, derived for an observer to object distance of
10\,pc, no conversion is required to derive absolute magnitudes and
therefore intrinsic colours.

We have integrated, the fluxes, using photon-counting and calibrated
using a Vega spectrum for magnitudes in the optical and near-IR
regimes. The filter bandpasses selected are, the optical system of
\cite{bessell_1998} for \textit{UBVRI} and the \textit{CIT} system of
\cite{elias_1982,stephens_2004} for \textit{JHK}, with the required
shift of $-0.015$\,$\mu$m as prescribed by \cite{stephens_2004}.

We have also derived magnitudes in the mid-IR range using the IRAC and
MIPS systems of the \textit{Spitzer} space telescope. These magnitudes
were derived using a conversion of flux to Data Number (DN) and
calibrated using zero points derived from the zero magnitude fluxes
published in the IRAC
handbook\footnote{http://ssc.spitzer.caltech.edu/documents/som/som8.0.irac.pdf}
and the MIPS instrument calibration
website\footnote{http://ssc.spitzer.caltech.edu/mips/calib/}.

We have chosen these specific filter due to their ubiquitous use and
suitability for the derivation of the key stellar parameters of age,
mass and, for populations, disc fractions. Therefore, by studying the
changes of magnitudes (and colours) in these photometric systems we
can test the predicted effects on these derived parameters caused by
changes in the input parameters of our model grid. The optical
magnitudes \textit{VI} are used to explore age-related effects of
varying the parameter space. Flux in the \textit{VI} bands is only
minimally affected by accretion flux \citep{gullbring_1998} and disc
thermal flux \citep{hartmann_1998} and, additionally, in a \textit{V,
  V$-$I} CMD, the reddening vector lies parallel to the pre-MS,
minimising any age effect of extinction uncertainty, \citep[for a full
discussion see][]{mayne_2007,mayne_2008}. The near-IR passbands of
\textit{JHK} are most often used to derive stellar masses. This is as
for pre-MS objects the reddening vector, in for instance, a \textit{J,
  J$-$K} CMD, is almost perpendicular to the isochrones. The direction
of this vector acts to minimise the mass effect of extinction
uncertainties \citep[the \textit{CIT} systems was chosen to
match][]{chabrier_2000}. Finally, as is now well documented the
\textit{Spitzer} IRAC passbands provide the best data with which to
unambiguously separate naked and star-disc systems
\citep{luhman_2005a}. In addition, at longer wavelengths, the MIPS
instrument provides sensitivity to disc systems at much larger radii
(or debris discs).

Once the model grid was completed several checks were performed to
verify the consistency of our results. For each individual model these
checks were passed to our satisfaction before publication. Some
problems remain, and these are explained in Appendix
\ref{consistency}.

\section{Parameter space}
\label{par_space}

This section details the range of each of the input parameters we have
varied and, where possible, gives justification for the selected
ranges from published observations. The simulations in this paper
cover variations in the stellar mass, stellar age, stellar rotation
rate, accretion rate, the areal coverage of the accretion stream, disc
mass fraction and the system inclination. A summary of the values
adopted for each input variable is shown in Table
\ref{par_space_table}.

\subsection{Mass} 
\label{par_mass}

Representative masses within the BD regime were chosen as follows:
0.01, 0.02, 0.03, 0.04, 0.05, 0.06, 0.07 \& 0.08\,$M_{\odot}$.

\subsection{Age}
\label{par_age}

Typical disc lifetimes for solar type stars are of order 10\,Myrs
\citep{haisch_2001}. Therefore, we have adopted input ages of
1 and 10\,Myrs for our model grid, to span the approximate range of
ages over which the discs influence will be important.

\subsection{Rotation rate}
\label{par_rotation}

Data for rotation rates, from periodic variability surveys, are widely
available for a range of different age clusters of TTS. However, fewer
studies exist on the rotation rates of BD mass objects. Rotation
period data for $\sigma$ Ori, at an age of $\approx$\,3 Myrs
\citep{mayne_2008}, was studied by \cite{scholz_2004}, where periods
are found over the range 5.78$-$74.4 hours ($\approx$\,0.24$-$3.1
days) for BD mass objects. \cite{scholz_2005} study rotation period
data for stars in the vicinity of $\epsilon$ Ori, with an assumed age
of $\approx$\,3 Myrs \citep{osorio_2002} and the ONC at an age of
$\approx$\,2 Myrs \citep{mayne_2008}. Rotation periods, from
photometric variability, in the range 4.7$-$87.6 hours
($\approx$\,0.2$-$3.65 days) for BD mass stars are
found. \cite{joergens_2003} study the rotational periods of BD (and
very low mass stars) in the Chameleon I region. This region is
$\le$\,1 Myrs old, and the authors find rotation periods of 2.19,
3.376 and 3.21 days for their BD targets.

In some cases the periodic variability is irregular and assumed to
come from active accretion hot spots on the BD surface \citep[see][for
a discussion of variability causes]{bouvier_1995,herbst_2007},
indicative of active accretion. All the studies mentioned infer a disc
locking mechanism. Furthermore, \cite{scholz_2004} and
\cite{scholz_2005} find evidence for a mass $\propto$ period
relationship extending into the BD regime. Additionally,
\cite{joergens_2003} propose a shorter lifetime of $\approx$\,5 Myrs for
BD discs, inferred from a shorter derived disc-locking
timescale. However, as discussed in the previous studies, an imperfect
disc locking mechanisms is also hypothesised as responsible for the
less significant loss in angular momentum out to ages of 10 Myrs, for
BD discs. The data on BD rotation rates, disc presence and disc
locking are summarised and discussed in \cite{herbst_2007}.

Therefore, to create a set of useful models to help contextualise the
observational constraints for study of disc locking mechanisms, we
must adopt a realistic and bounding range of rotation rates. For our
model grid, and associated age range ($<$10 Myrs), we have selected
0.5 and 5 days. With the limits set at at the approximate median of
faster rotators and the edge of the slower rotators.

\subsection{Areal Coverage}
\label{par_cov}

As discussed in Section \ref{par_rotation} evidence for irregular
periodic variability has been found in BDs with detections of
associated stellar discs. This is construed as evidence for accretion
hot spots formed as magnetically channeled material hits the stellar
surface \citep[see discussion in][]{bouvier_1995,herbst_2007}. The
irregularity is thought to be caused by changes in the magnetospheric
structure and accretion rate \citep{bouvier_1995}. For our model we
have assumed that disc material is disrupted at the co-rotation radius
and channeled onto the star in the form of accretion hot spots with a
characteristic temperature. Therefore, to calculate the characteristic
temperature and the resulting blackbody accretion flux we must adopt
an accretion rate and areal coverage of the accretion stream.

Little observational evidence can be found for approximate sizes of
accretion hot spots due to their more transient nature and often
smaller coverages, when compared to cooler or `plage' spots
\citep{herbst_2007}. \cite{bouvier_1995} modelled the
size of the cool spots on solar-type stars for a selection of
periodically variable candidates. They found projections of cooler
spots, onto the stellar disc, of a few to $\sim$60\%.
\cite{bouvier_1995} also found projected sizes, onto the
stellar disc, of typically a few \% to around 10\% for hot spots.
\cite{bertout_1996} used observations of YY Orionis monitoring
flux amplitude variations as a function of wavelength to derive a
probable hot spot area of around 10\%. The spot temperature was also
modelled for YY Orionis in \cite{bertout_1996}, resulting in a
best fitting areal coverage of 11\%. Therefore, to bound the probable
areal coverage range of the accretion hot spots we have adopted areal
coverages of 1 and 10\%.

\subsection{Accretion Rate}
\label{par_accn}

Accretion rates derived for pre-MS stars are of order \logmdot = $-6$
to $-11$ \citep{natta_2006}, with the largest accretion rates found in
so-called FU Orionis type objects. For the more typical accretion
rates \citep[$\dot{M}=$10$^{-11}$ to 10$^{-8.9} M_{\odot}yr^{-1}$, for
TTS,][]{dahm_2009}, several studies have now suggested that the
accretion rate is strongly correlated with the mass of the central
star. This relationship was first suggested by \cite{muzerolle_2003}
using various accretion diagnostics. Later, \cite{muzerolle_2005}
derived a relationship of approximately $\dot{M}\propto
M_*^2$. Further evidence was put forward by \cite{natta_2004}, where
accretion rates as low as 5$\times$ 10$^{-12} {\rm M}_{\odot} {\rm
  yr}^{-1}$ were found for BDs. More recently, even lower
accretion rates of $\approx$\,10$^{-13} {\rm M}_{\odot} {\rm
  yr}^{-1}$, have been derived for BDs by \cite{herczeg_2009}.
Further support for a dependence of accretion rate on stellar mass was
apparent in the significantly more homogeneous dataset of
\cite{natta_2006}. \cite{natta_2006} analysed a set of accretion rates
and masses derived for BDs in $\rho$ Ophiuchi and compared these
results to stars in Taurus. They found that the accretion rate scales
with central object mass into the BD regime, although with significant
scatter.

As the relationship $\dot{M}\propto M_*^2$ predicts lower accretion
rates for BD mass objects it is essential that we model systems at
higher accretion rates, which may have been missed in current
observational studies. Therefore, we have adopted accretion rates of
\logmdot = $-6$, $-7$, $-8$, $-9$, $-10$, $-11$ \& $-12$.

\subsection{Disc Mass}
\label{par_mdisc}

Previously studies modeling BD discs have adopted a range of disc mass
fractions, for instance \citep{walker_2004} use 0.1, 0.01 and
0.001$M_*$. \cite{wood_2002} fitted observed spectra with modelled SEDs
to derive a disc mass of 0.003$M_*$  for HH~30 IRS. Subsequent
derivations of disc masses have converged to within an order of
magnitude, with the following specific results: 0.03$M_*$
\citep[$\rho$ Ophiuchi,][]{natta_2002}, 0.055$M_*$ \citep[GM
Aurigae,][]{rice_2003}, 0.03$M_*$ \citep[GY 5, GY 11, and GY
310,][]{mohanty_2004} and 0.022$M_*$\citep[2MASS
J04442713+2512164,][]{bouy_2008}. As the derived disc masses all have
a similar order of magnitude we have adopted $M_{\rm disc}\approx
$\,0.01$M_*$. As changes in disc masses are expected to change the
resulting SED less than perhaps, accretion rate for example, we have
not varied the disc mass for this study. The results of simulations
varying this parameter will be published in a future paper.

\subsection{Inclination}
\label{par_inc}

Discs around BDs exhibit increased flaring, due to the reduced
surface gravity in the disc \citep{walker_2004}. This increased
flaring, and therefore larger scaleheight of the disc results in
obscuration on the star at lower inclinations, when compared to higher
mass stars and their circumstellar discs. As has been shown in
\cite{walker_2004} effects caused by variations in the system
inclination angle are much more significant for BDD systems, again
compared to their higher mass analogues. Therefore, we have simulated
ten observer to system inclination angles spaced evenly in $\cos i$
space, namely, 0, 27, 39, 48, 56, 64, 71, 77, 84 and 90$^{\circ}$.

A final list of all varied parameters and their values can be seen
in Table \ref{par_space_table}.

\begin{table*}
\begin{tabular}{|l|l|}
\hline
Input parameter&Values (\# of values)\\
\hline
Mass ($M_{\odot}$)&0.01, 0.02, 0.03, 0.04, 0.05, 0.06, 0.07 \& 0.08 (8)\\
Age (Myr)&1 \& 10 (2)\\
Rotation period (days)&0.5 \& 5 (2)\\
Areal coverage (of $\dot{M}$, \%)&1 \& 10 (2)\\
Accretion rate (log$(\frac{\dot{M}}{M_{\odot}} yr^{-1})$)&$-6$, $-7$, $-8$, $-9$, $-10$,
$-11$ \& $-12$ (7)\\
Disc mass ($M_*$)&0.01 (1)\\
Surface density profile & $r^{-1}$ (1) \\
Inclination ($^{\circ}$)&0, 27, 39, 48, 56, 64, 71, 77, 84 \& 90
(10)\\
\hline
\end{tabular}
\caption{List of all varied input parameters. Resulting in a total
  number of models of 448 (plus 40 models without radiative transfer
  simulations for the naked BDs) and 4480 SEDs (plus 40 for naked
  BDs). \label{par_space_table}}
\end{table*}

\section{Result and Analysis}
\label{results}

In this section we first discuss the physical structure, both density
and temperature, of the BDD disc systems (Section \ref{disc_struct})
across our parameter space. Then we discuss the resulting simulated
observations in Section \ref{observables}. We present analysis in
terms of the impacts of the disc structure on the SEDs and colours and
magnitudes. Then in Section \ref{isochrones} we examine the
reliability of age, mass and disc fraction derivation when applied to
our model grid. In particular we discuss selection effects causing
higher accreting systems to be unlikely to be classified as BDD
systems. Essentially, despite not intrinsically including a dependence
of accretion rate on stellar mass in our grid, we show that current
observational techniques and theoretical models, applied to the grid
,would result in a relationship of this type being derived.

\subsection{Disc Structure}
\label{disc_struct}

The disc surface density is conserved in the initial density
structure, $\Sigma ^{\beta - \alpha}$. The initial scaleheight ($h$)
and density ($\rho$) follow $h=h_0(r/R_*)^{\beta}$$\rho
=\rho_0\frac{R_*}{r}exp(-\frac{1}{2}[z/h(r)]^2)$ respectively., where
$r$ and $z$ are the radial and vertical coordinates. The initial
values of $\alpha$ and $\beta$ were 2.1 and 1.1 respectively. The
values for $\alpha$ and $\beta$ were chosen to optimise resolution of
the vertically evolving disc, but minor variations are largely
inconsequential as the systems evolves from this state. In our models
we have, however, placed the inner disc edge at the co-rotation
radius, as opposed to the dust destruction radius used in
\cite{walker_2004}. The initial disc scaleheight at 100\,AU, $h(100)$,
was set to 25\,AU. As the simulation used vertical hydrostatic
equilibrium and dust sublimation, both the disc scaleheight and inner
edge location then evolved in the systems dependent on the input
parameters. In this section we discuss the structure of the discs in
terms of these two generated characteristics, i.e the disc scaleheight
and inner edge location.

\subsubsection{Disc Flaring}
\label{disc_flaring}

\begin{figure*}
\begin{center}
  \subfigure[]{\includegraphics[scale=0.4,angle=0]{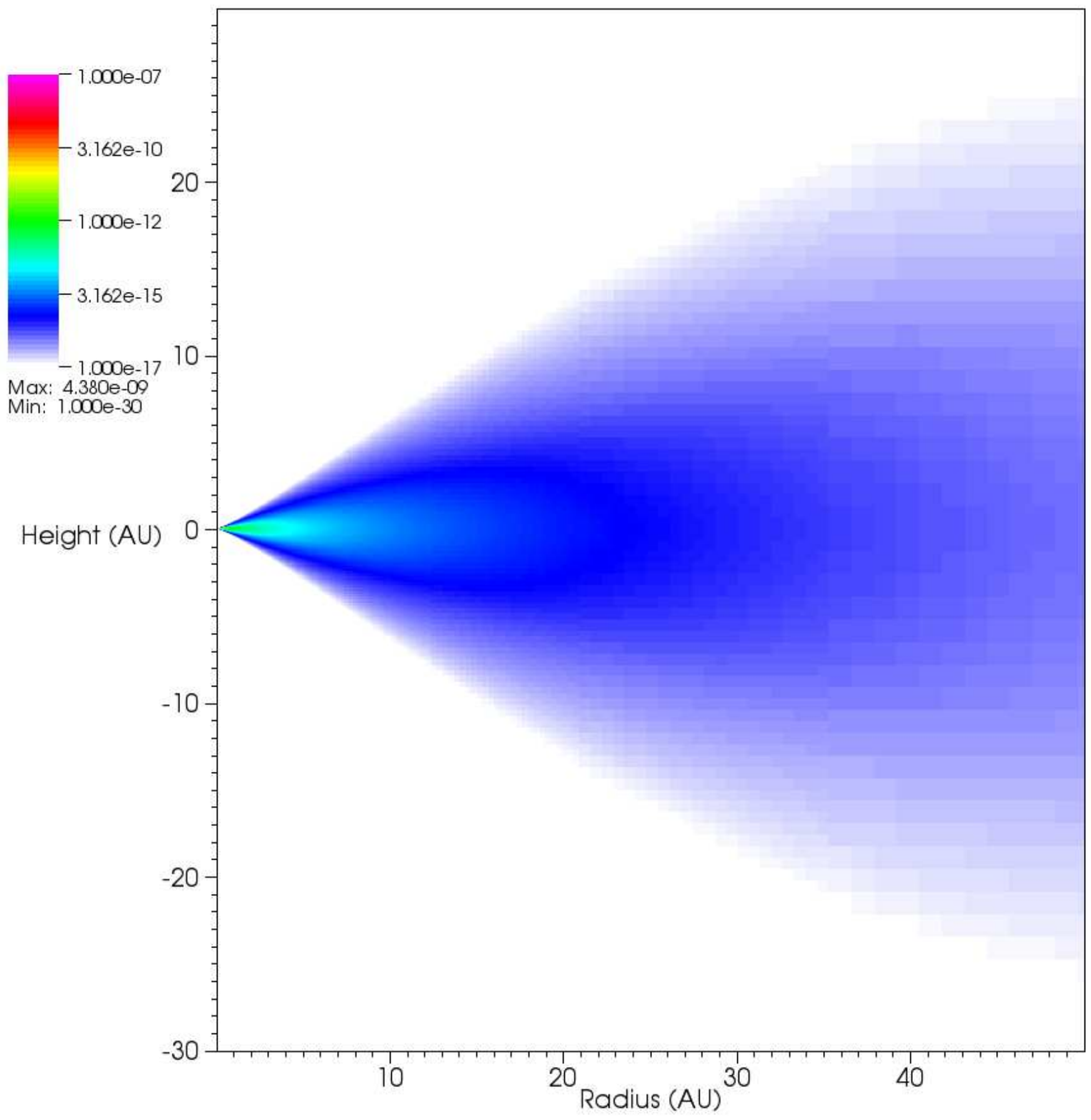}\label{flare_-12_183}}
  \subfigure[]{\includegraphics[scale=0.4,angle=0]{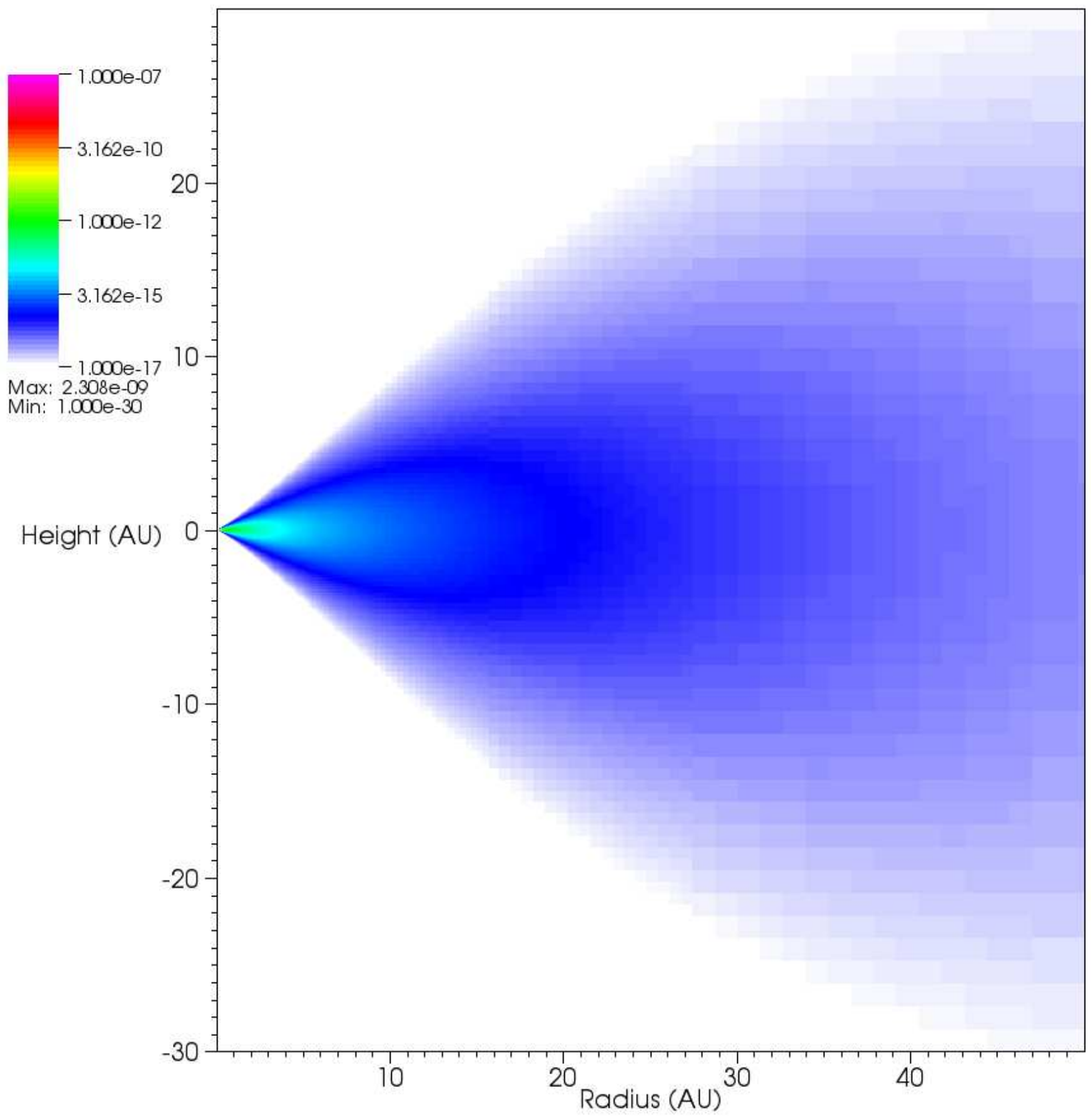}\label{flare_-7_193}}
\end{center}
 \caption{Figure showing the density structure ($\log \rho$) of the
   BDD system with $M_*=0.04M_{\odot}$, Age=1 Myrs, $\tau$=5, areal
   coverage=10\% and accretion rate of (a) \logmdot=$-12$ and (b) \logmdot = $-7$.}
\end{figure*}

We have previously tested the results of the {\sc torus} code against
that used by Walker and co-workers, using a CTTS disc model and
simultaneously solving for radiative and hydrostatic equilibrium (but
not employing dust sublimation). These tests showed excellent
agreement in density and temperature structure, as well as in the
resultant SEDs. The results of these tests were presented by
\cite{walker_2006}. It is therefore unsurprising that at negligible
mass-accretion rates our disc structures are very similar to those
present in \cite{walker_2004}. Typical discs around CTTS stars have
scaleheights at 100\,AU of between $h$(100)$=$10 to 20\,AU, whereas
for BDD systems, $h$(100)=20 to 60\,AU (for 0.08 and 0.01 $M_{\odot}$
respectively). As the accretion rate increases the flux levels of the
central star increase and lead to heating of the disc which in turn
leads to vertical expansion. We found that levels of vertical flaring
increased only marginally with accretion rate. Significant
differences, more than $>$5\,AU increase in $h(50)$, in the vertical
structure were not apparent until the high accretion rates of \logmdot
= $-7$ and $-6$. Figures \ref{flare_-12_183} and \ref{flare_-7_193}
show the density structure ($\log \rho$) in the disc from radial
distances of 0 to 50\,AU for example systems ($M_*=0.04M_{\odot}$,
Age=1\,Myrs, $\tau$=5\,d and areal coverage=10\%), with accretion
rates of \logmdot = $-12$ and $-7$, respectively.

\cite{walker_2004} state that the degree of disc flaring depends on
the disc temperature structure and the mass of the central star, with
the disc scaleheight $h\propto (T_{\rm disc}/M_*)^{1/2}$
\citep{shakura_1973}. Recently however \cite{ercolano_2009} proposed
that the flaring varied in the opposite sense with stellar mass,
i.e. $h\propto M_*$. This suggestion was based on evidence from
\cite{allers_2006}, where SEDs for 17 systems in the mass range
6$M_{\rm Jup}<M_*<$350$M_{\rm Jup}$ were fit with flared or flat disc
models. In general, \cite{allers_2006} find that lower mass objects
achieve better fits with the flat disc models and higher mass objects
with the flared discs.

The results of \cite{allers_2006} show that above a mass of 50$M_{\rm
  Jup}$ all objects (6/17) are better fit with flared discs. Whilst at
masses below 50$M_{\rm Jup}$ only one object is better fit by the
flared disc model, with the remaining objects (10/17) better fit with
flat models. Whether this result is statistically significant enough
to assert a $h\propto M_*$ is doubtful as the fitting process contains
(presumably) two fixed scaleheight distributions. Therefore, for our
study we continue to assume that our flared BDD systems will have
larger characteristic scaleheights than typical CTTS systems.

A comparison of Figures \ref{flare_-12_183} and \ref{flare_-7_193}
shows an increase in the scaleheight at 50\,AU of $>$5 AU, as the
accretion rate moves from \logmdot =$-12$ to $-7$. However, despite
this small change with high levels of accretion our grid shows
scaleheights comparable to the work of \cite{walker_2004} and as such
result in similar consequences for the SEDs and photometric
magnitudes. The effects of this flaring and the increase in flaring
for very high accretion levels on SEDs and photometric magnitudes are
discussed in Sections \ref{observables} and \ref{isochrones}
respectively.

\subsubsection{Inner edge of the dust disc: location}
\label{inner_edge}

The inner edge of the gas disc is fixed at the co-rotation radius, at
which point the gas threads onto the magnetic field and follows the
field lines in a funnel flow towards the protostellar surface. We make
the assumption that the dust (should it exist at the co-rotation
radius) is destroyed in the funnel flows. This is a reasonable
assumption from both theoretical and observational perspectives:
Radiative-transfer models indicate the temperature in the funnel flows
may be very much greater than the dust sublimation temperature
\citep[$>10$\,kK,][]{muzerolle_2003}, while dusty funnel flows are
likely to be optically thick in the visible regime, which would lead
to substantial photometric variability as the funnels transit the
photosphere--an effect that is unobserved.

Our models include a sophisticated treatment of dust sublimation
(described in Section \ref{dust_edge}).  As the flux levels of the
central protostar increase with increasing accretion rates the flux
incident on the inner edge increases and leads to increasing erosion
of the inner edge of the dust disc.

As the inner edge moves, its temperature is expected to change. This
has been predicted to lead to a correlation of inner edge position and
IR excess \citep{meyer_1997}. This may act to bias surveys correlating
rotation rates with IR excess to search for evidence of
disc-locking. However, the flux from the inner edge will usually peak
between 2 and 3\,$\mu$m \citep{dullemond_2001}, given that the dust
sublimation temperature peaks at $\approx$\,1400\,K for canonical
densities. This means that for models where dust is significantly
sublimated, the inner edge will {\em usually} have a temperature of
$\approx$\,1400\,K and this correlation of disc position and
temperature will be lost (although see below).

Equation \ref{inner_eq} shows that as the rotation period of the
protostar increases the co-rotation radius decreases. This will result
in, initially, shorter period systems having closer and hotter inner
edges than their longer period counterparts. In addition, if the
accretion rate is increased in these systems the incident flux on the
inner wall will increase leading to a rise in the temperature of the
inner edge.  At some point the dust sublimation temperature may be
reached, leading to a change in radial position of the wall. In
addition, the temperature of the inner edge will then tend to the dust
sublimation temperature.

A further complexity arises when one considers that the density of
material in the disc falls with increasing radius from the star
($\rho(r) \propto r^{-\alpha}$), and that the dust sublimation
temperature is density dependent \citep{pollack_1994}.  Therefore, for
systems where the inner edge has been eroded significantly from the
co-rotation radius, the temperature of the inner edge will fall
systematically as the inner dust disc radius increases.

\begin{figure*}
\includegraphics[scale=0.5,angle=90]{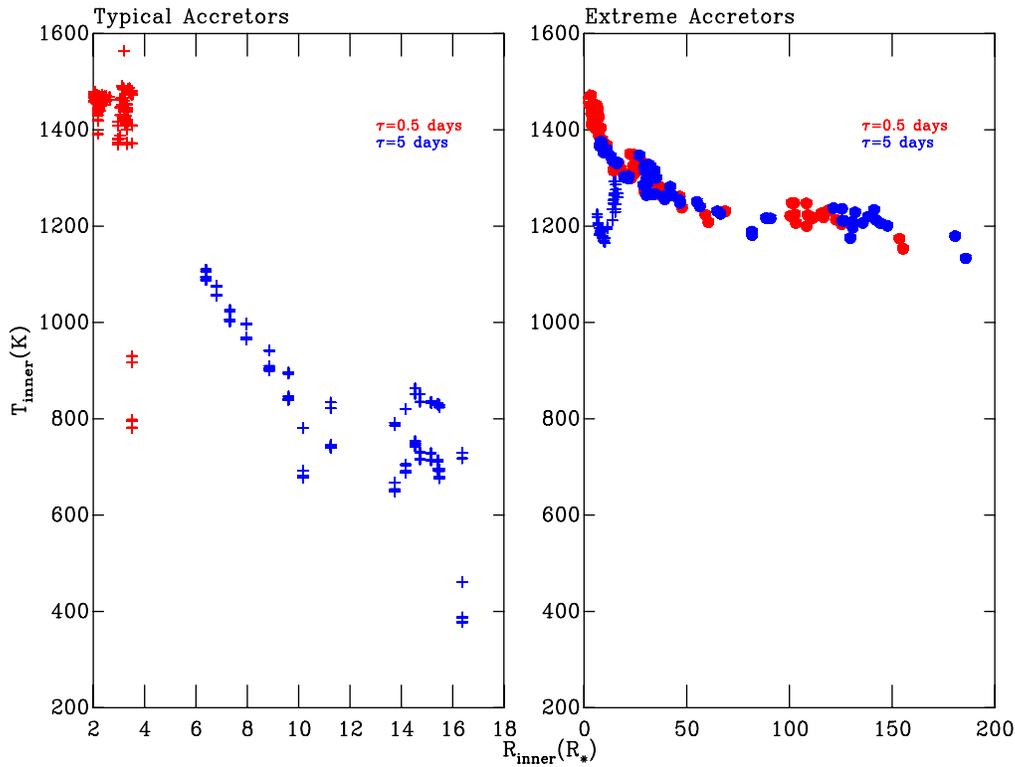}
\caption{Figure showing data for the inner edge location ($R_{\rm
    inner}$, $R_*$). The typical accretors are shown in the left panel
  and the extreme accretors in the right panel (see text for
  explanation). Both panels separate the systems by period, with those
  systems with rotation periods of 0.5 and 5 days shown as red and
  blue symbols respectively. In addition, systems where $\Delta R_{\rm
    inner} < 1R_*$ are plotted as crosses and $\Delta R_{\rm inner} >
  1R_*$ as filled circles (this is only achieved by some systems
  classed as extreme accretors).\label{inner_temp}}
\end{figure*}

For further analysis, and throughout the rest of this paper, we will
separate the systems into two groups; those systems with
\logmdot$\leq-$9 and \logmdot$>-$9, classed as typical and extreme
accretors. This distinction is based on the point at which the systems
undergo both significant sublimation ($\Delta R_{\rm inner}>1R_*$) and
have an accretion luminosity that is comparable to the photospheric
luminosity (see Section~\ref{acc_dominance}). In order to examine the
effect of dust sublimation on our models we have measured the radius
of the inner edge of the dust disc by integrating from the centre
along the midplane until an optical depth of 2/3 (at 5500\AA) is
reached. Figure \ref{inner_temp} shows the radial density distribution
of the disc.  In this case, the final inner edge radius, and therefore
temperature, is no longer strongly correlated with the rotation
rate. Figure \ref{inner_temp} shows the final inner edge location
($R_{\rm inner}$) against the temperature of the inner wall ($T_{\rm
  inner}$). The left panel shows the systems designated as typical
accretors and the right panel those with extreme accretion rates.  For
both panels the systems with rotation periods of 0.5 and 5 days are
plotted in red and blue respectively. Those systems where the change
in inner radius was greater than 1$R_*$, classed as significantly
sublimated, are plotted as crosses.

The left panel of Figure \ref{inner_temp} shows that, taken
as a whole, our models with typical accretion rates show a clear
correlation between the temperature at the inner edge and the radius
to this boundary. This agrees with the work of \cite{meyer_1997} where
this correlation is found for there $R_{\rm inner}=$1$-$12$R_*$ and
\logmdot = $-9$ to $-5$, for flat disc models.  \cite{meyer_1997} use
this correlation, and the derivation of IR magnitudes, to predict a
relationship between IR excess and radius to the inner wall. Our data
indicate that this correlation, for typically accreting systems will
translate into a correlation between rotation rate and IR excess. This
could have important implications for studies of disc-locking where
disc presence is examined as a function of rotation rates, provide an
intrinsic bias. In practice however, this correlation is weak (in fact
unobservable) in our data due to the combined effects of the inner
disc wall shape, inclination and flaring effects. This is discussed in
more detail in Section \ref{isochrones}.

Figure \ref{inner_temp} shows that for those systems where significant
disc erosion ($\Delta R_{\rm inner} > 1 R_*$) has occurred the
resulting temperature of the inner edge is weakly correlated with the
inner edge radius, but, critically, not correlated with rotation rate.
The inner edge temperatures of the remaining systems for the extreme
accretors are slightly anti-correlated with the radius to the inner
edge. For the systems with typical acccretion rates, and longer
periods, Figure \ref{inner_temp} shows there is again a weak
correlation between the radius to, and the temperature of the inner
edge. For the shorter period models with typical accretion rates there
is no clear correlation between the temperature at the inner edge and
radius to this edge. 

\subsubsection{Inner edge of the dust disc: shape}
\label{inner_edge_shape}

\begin{figure*}
  \centering
  \subfigure[]{\includegraphics[scale=0.27,angle=0]{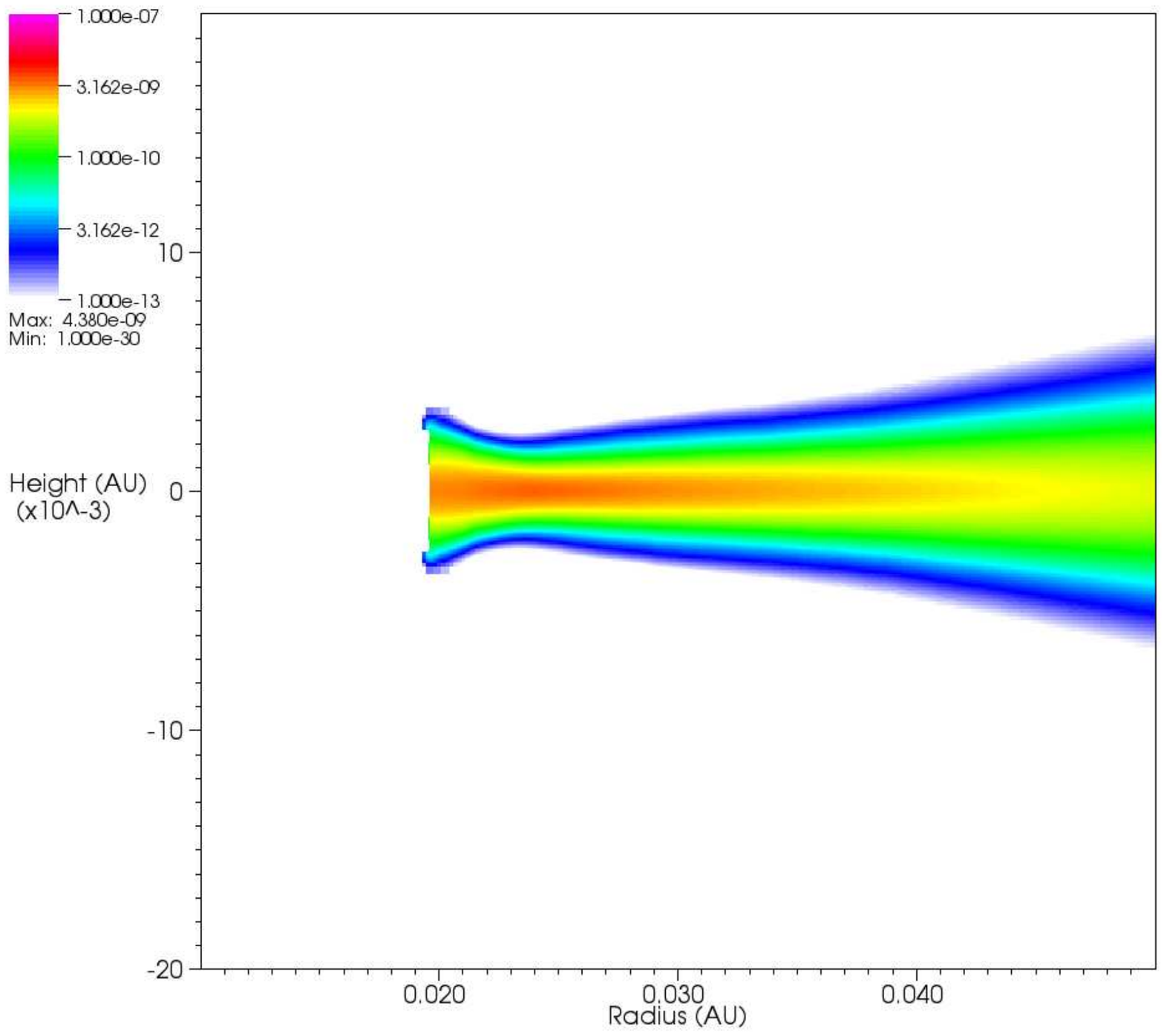}\label{no_sub_183_a}}
  \subfigure[]{\includegraphics[scale=0.27,angle=0]{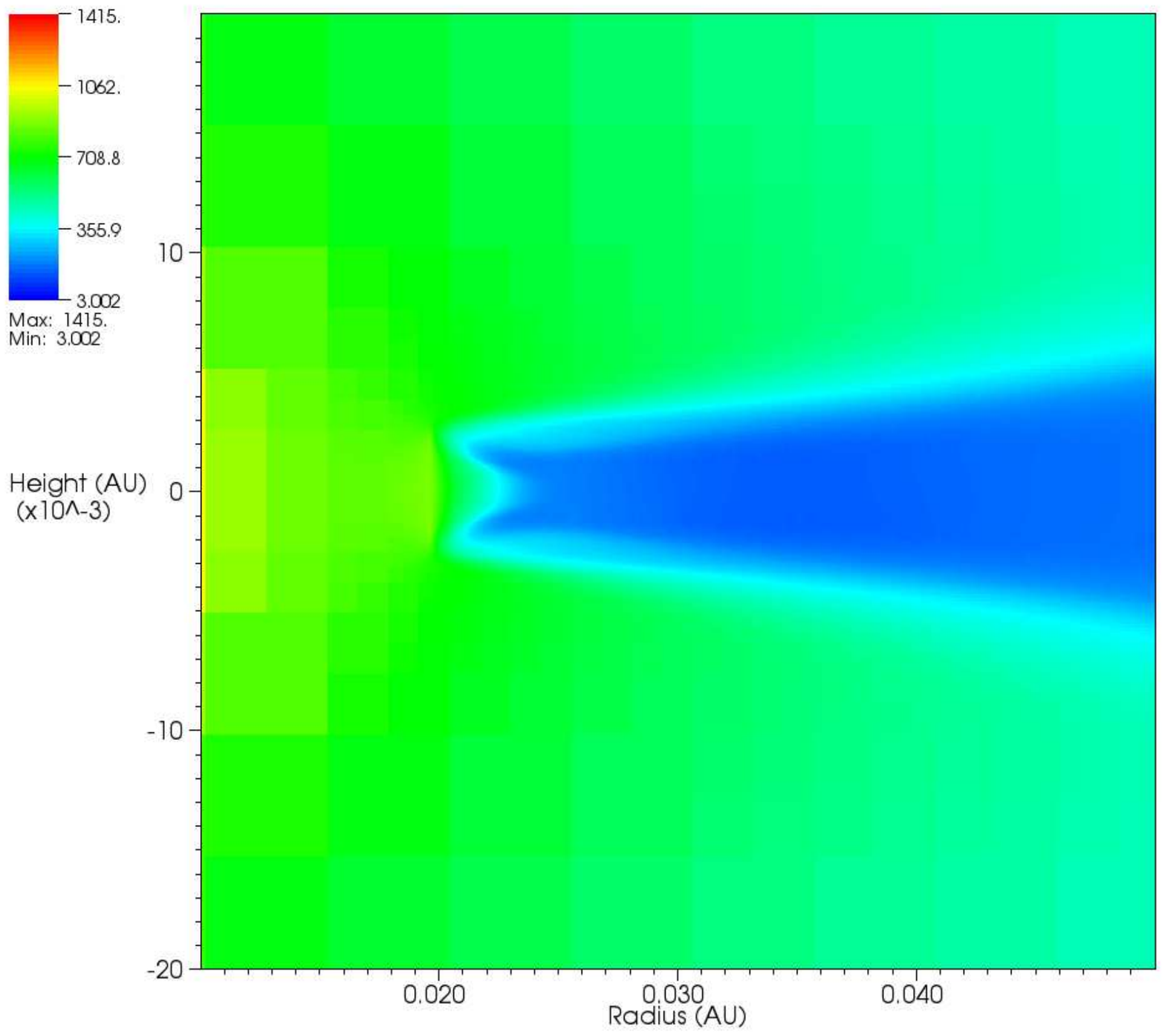}\label{no_sub_183_b}}
  \subfigure[]{\includegraphics[scale=0.27,angle=0]{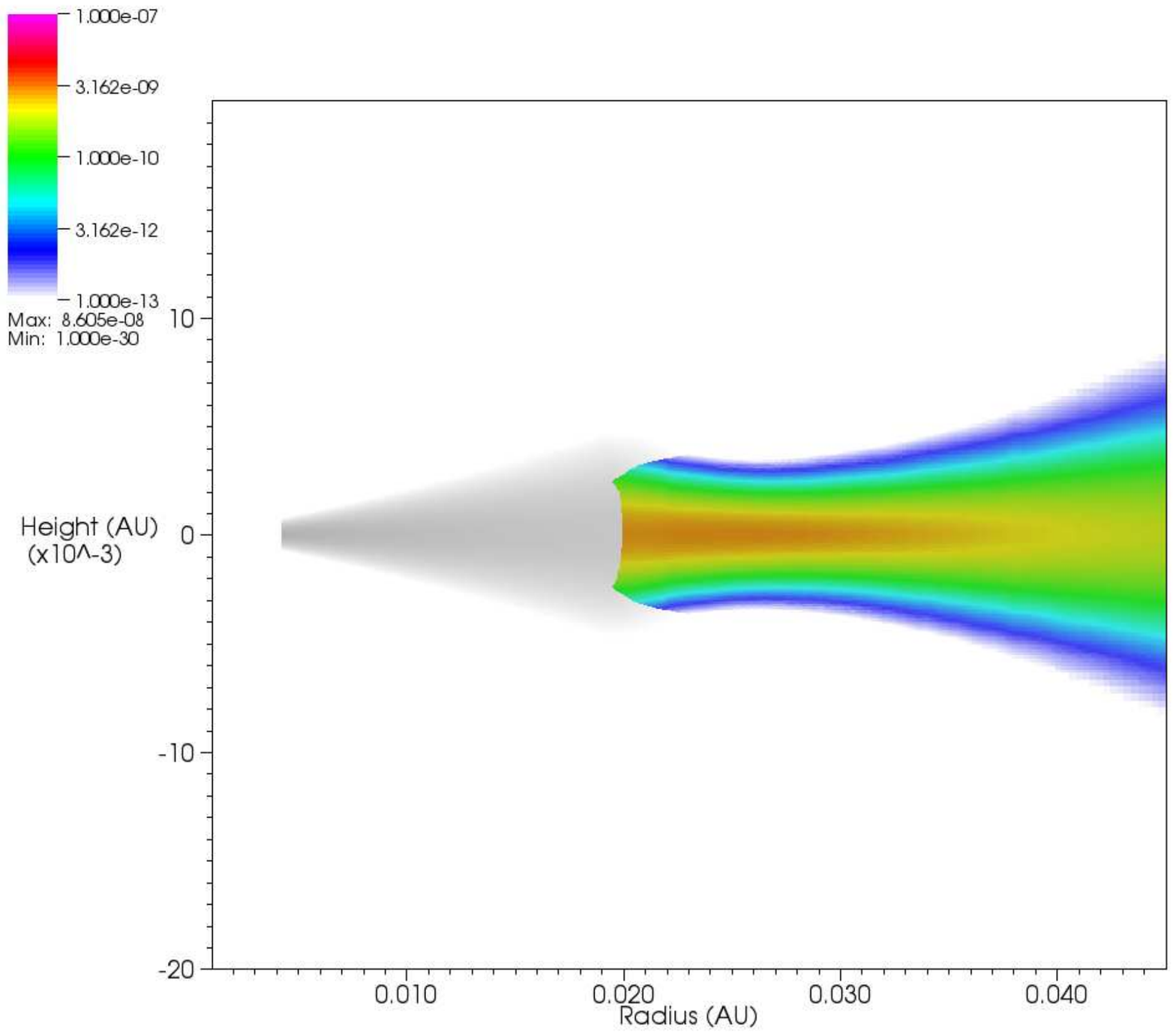}\label{sub_179_a}}
  \subfigure[]{\includegraphics[scale=0.27,angle=0]{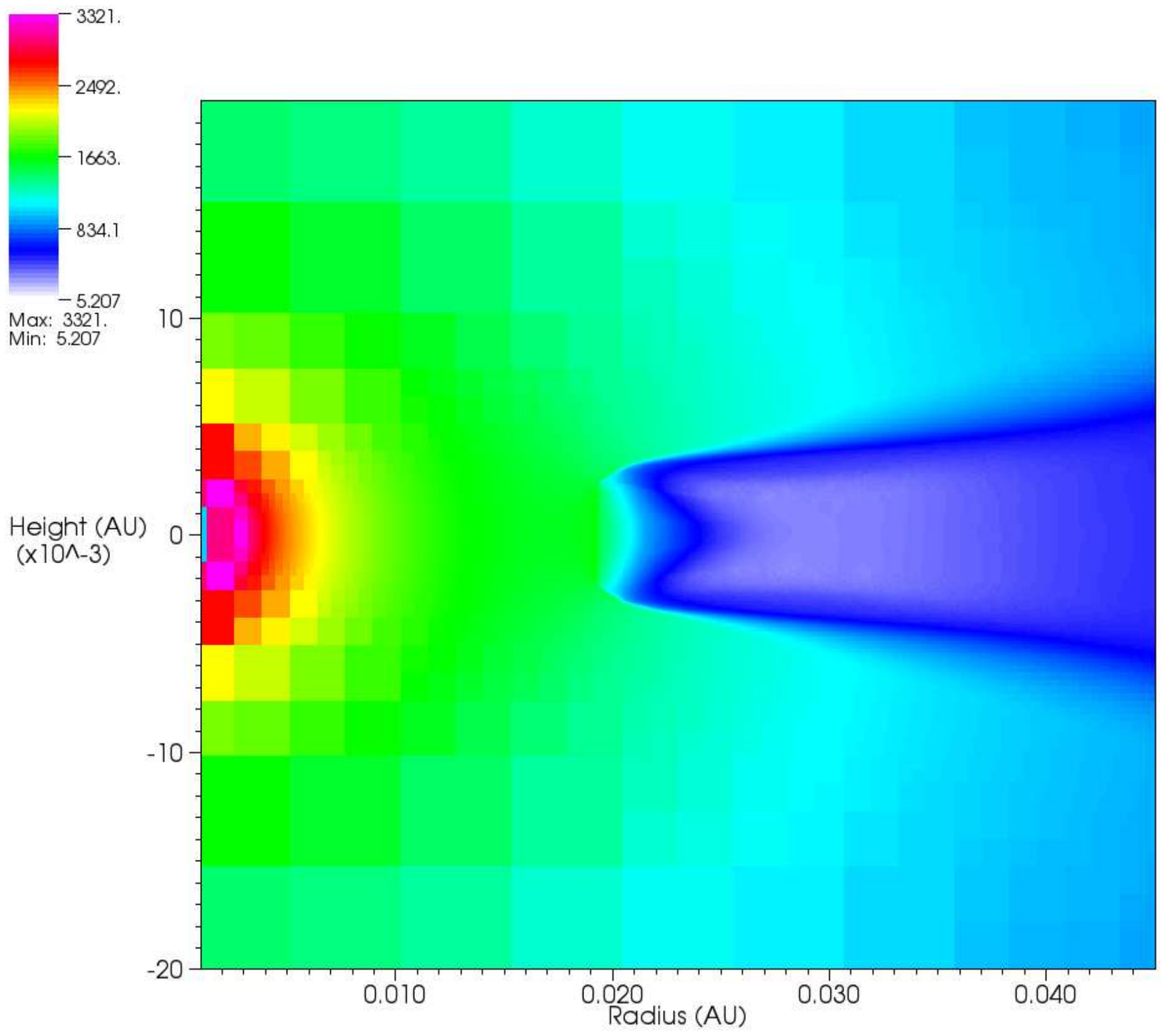}\label{sub_179_b}}
  \subfigure[]{\includegraphics[scale=0.27,angle=0]{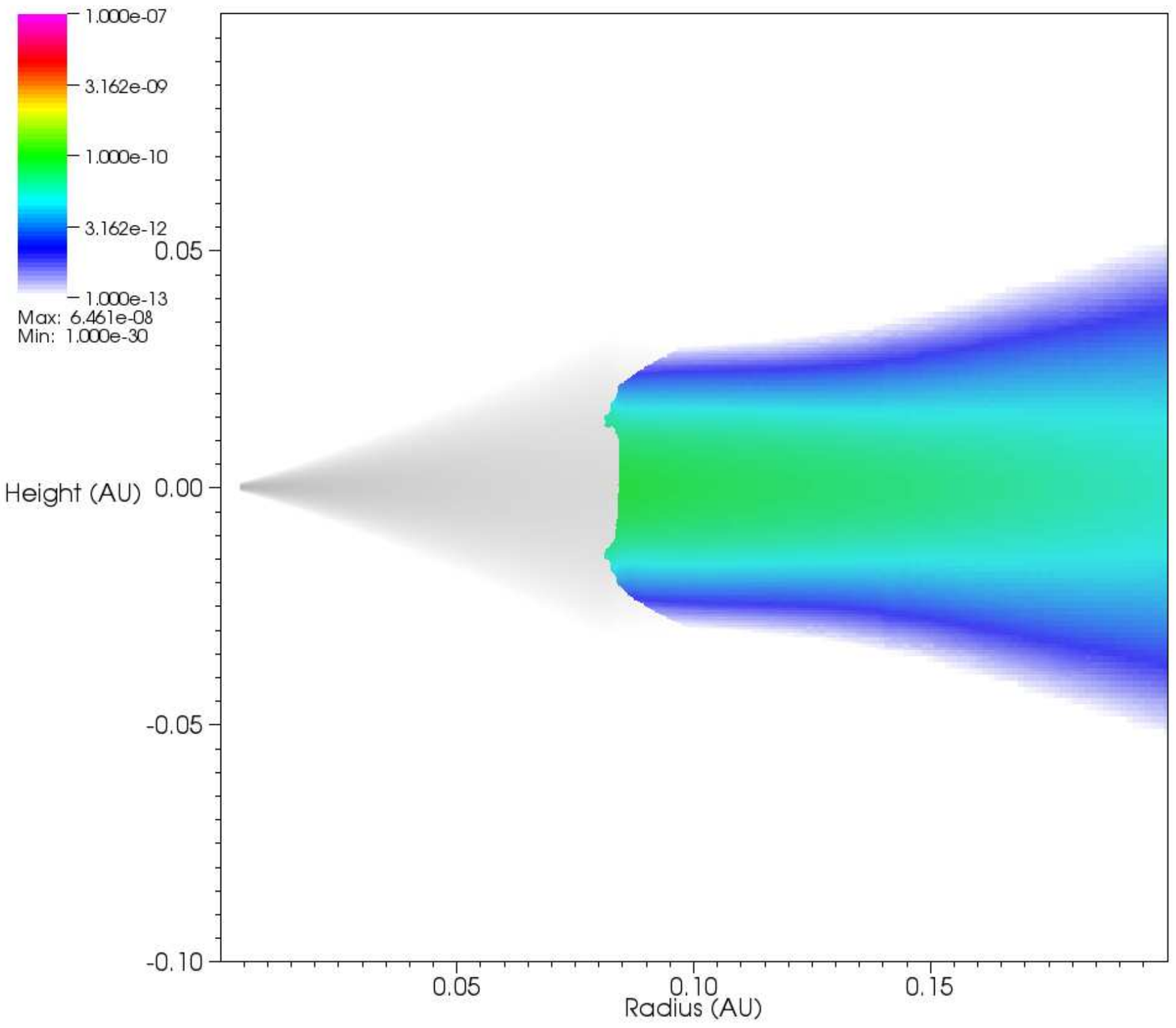}\label{sub_181_a}}
  \subfigure[]{\includegraphics[scale=0.27,angle=0]{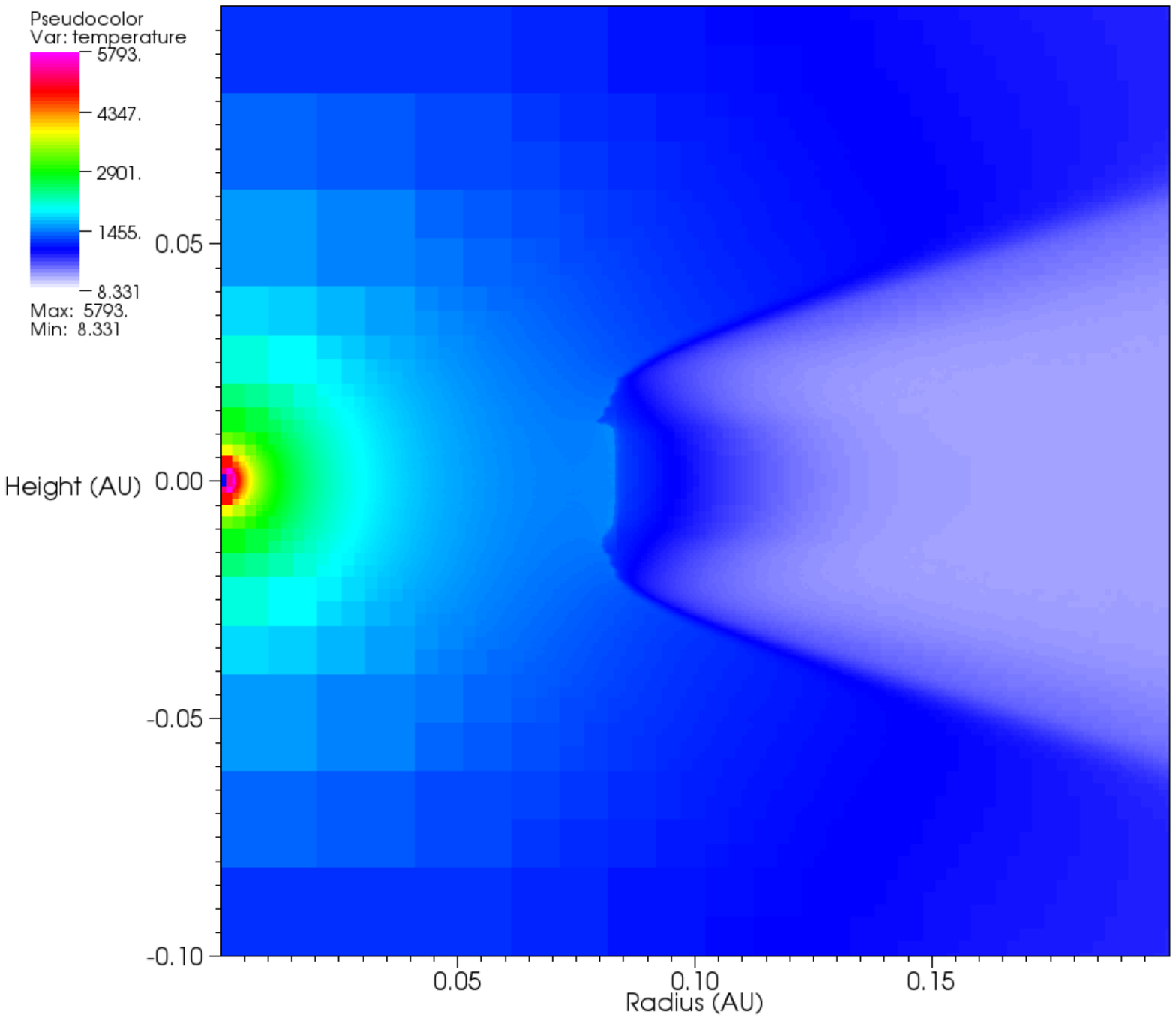}\label{sub_181_b}}

  \caption{Figure showing both the initial (grey scale) final density
    ($\log \rho$) where dust is present (colour scale, left-hand
    panels) and temperature (right-hand panels). System parameters
    are: $M_*$=0.04$\,M_{\odot}$, Age=1\,Myrs and areal coverage=10\%,
    $\tau$=5\,days and accretion rate \logmdot$=-12$ (a, b);
    $M_*$=0.04$M_{\odot}$, Age=1\,Myrs and areal coverage=10\%,
    $\tau$=0.5\,days and accretion rate \logmdot$=-7$ (c, d);
    $M_*$=0.04$M_{\odot}$, Age=1\,Myrs and areal coverage=10\%,
    $\tau$=0.5\,days and accretion rate \logmdot$=-6$ (e, f).}
\end{figure*}

The initial shape of the inner edge of the dust disc is a vertical
wall coincident with the co-rotation radius. In the previous section
we have shown that models with a negligible accretion rate do not
significantly sublimate the dust, and hence the edge remains
vertical. This inner edge is heated by direct radiation from the
protostar, and its scaleheight increases. Disc material behind the
inner rim is shielded from direct radiation and has a smaller
scaleheight, leading to the `puffed up' inner rim predicted by
\cite{dullemond_2001}. This effect is illustrated in
Figure~\ref{no_sub_183_a}, a model in which there is negligible dust
sublimation.

As the flux of the central object increases significant dust
sublimation occurs, leading to a change in the radial position of the
dusty inner edge, but also shaping it: The density drops rapidly away
from the midplane, and since the dust sublimation temperature also
falls the edges of the inner dust disc become curved--this the
mechanism described analytically by \cite{isella_2005}. We illustrate
this effect in Figure~\ref{sub_179_a}, which also shows a scaleheight
decrease behind the curved inner rim, over a significantly larger
radial scale than the previous model.

Finally the most extreme accretor (Figure~\ref{sub_181_a}) shows dust
being destroyed out to very large radii ($\sim 0.1$\, AU). A curved
rim is present, without any obvious decrease in scaleheight behind the
inner rim. The distance from the central object is such that the
vertical component of gravity is much diminished, and the disc has a
substantial scaleheight at the inner edge, meaning that it reprocesses
significant protostellar radiation leading to a high near-IR excess.

We note that a similar sequence is apparent across the grid for set
masses. However, the balance of the rotation period, and therefore
inner edge location, and age and areal coverage, therefore flux
levels, leads to changes in the accretion rate at which the dust
sublimation starts. However, in almost all cases the dust sublimation
does not become significant until at least \logmdot = $-9$ (as
discussed previously).

\subsection{Observable Consequences of Disc Structure and Accretion}
\label{observables}

The resulting converged disc structures, as discussed in Section
\ref{seds} are then used to create simulated SEDs and derive broadband
photometric magnitudes. In this section we discuss the effects of
accretion and disc presence on the simulated observations.

\subsubsection{Accretion dominance}
\label{acc_dominance}

As $T_{\rm acc} \to T_{\rm eff}$, disentangling the accretion and
stellar photospheric flux, in order to derive accretion rates becomes
difficult. This equivalence in accretion hot spot and stellar
photospheric temperature, for an accretion rate of \logmdot = $-9$,
occurs at a fractional coverages of 20 and 10\%, for rotation periods
of 5 and 0.5 days (with $M_*=0.04M_{\odot}$ and an age of 1
Myr). Additionally, higher accretion rates can produce enough flux to
`veil' the underlying photospheric features, and significantly change
the peak flux levels. Heavy veiling of atomic lines has been
extensively observed in CTTS systems \citep[e.g][]{kenyon_1995}.

\begin{figure*}
\includegraphics[scale=0.6,angle=90]{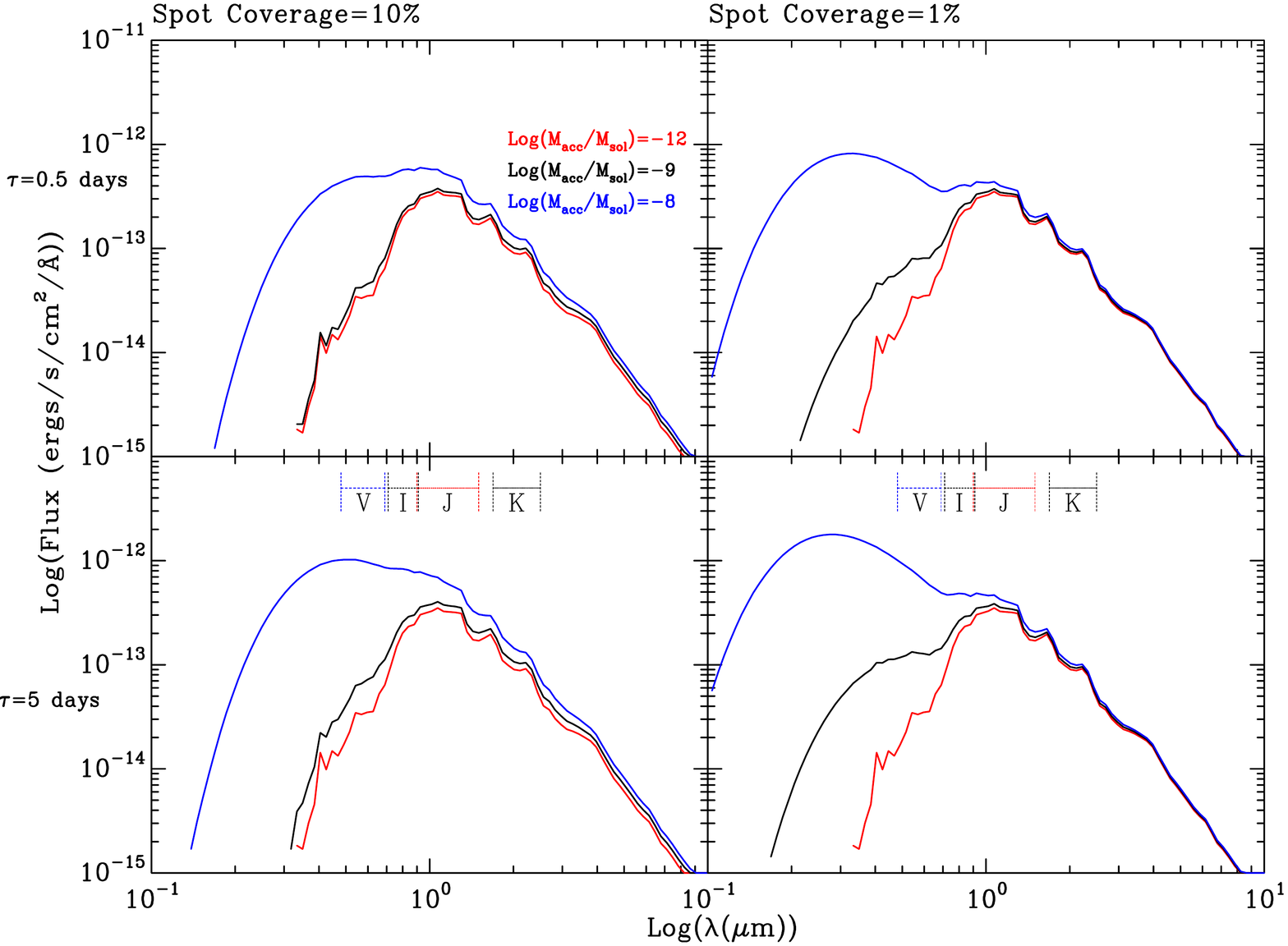}
\caption{Figure showing the photospheric flux (log(ergs$\rm{/s/cm}^2/{\rm
    \AA}$)) against $\lambda$ (log($\mu$m)) of a BD with
  $M=$0.04$M_{\odot}$, at 1 Myr. No disc is included, but blackbody
  fluxes from an accretion stream at the rates of \logmdot $=-8$, $-9$
  and $-12$ are shown as blue, black and red lines respectively. The
  bottom panels show accretion for a star rotating at 5 days, with the
  top panels showing that of 0.5 days. The left panels systems with an
  areal coverage of 10\% and the right panels has 1\%. The vertical
  and horizontal dashed lines (in the lower panels) denote the
  approximate sensitivity ranges of our chosen $V$, $I$, $J$ and $K$
  filters.\label{acc_dom}}
\end{figure*}

Figure \ref{acc_dom} shows the effect of increasing the accretion
blackbody flux (for increasing accretion rates) for a BD,
$M=$0.04$M_{\odot}$, at 1 Myr. Whilst the photon packets originating
from the star (both from $L_*$ and $L_{\rm acc}$) will be tagged as
stellar by {\sc torus}, we can separate these flux contributions
simply by observing the naked star system. The panels in Figure
\ref{acc_dom} show the flux from a naked system, with no treatment of
the disc. This enables us to view the effect of increasing accretion
rate on the photospheric flux in isolation. The accretion rates
included in all panels are \logmdot = $-8$, $-9$ and $-12$ (blue,
black and red lines respectively). The bottom panels show the systems
with a rotation period of 5 days and top panels for those with a
rotation period of 0.5 days. Given our assumption that accretion
occurs from the co-rotation radius, decreasing the rotational period
moves this accretion radius closer to the star, $R_{\rm inner}\propto
\tau^{2/3}$ (see Equation \ref{inner_eq}). As the accretion radius
moves further from the star the potential energy released by the
accreted material is increased. This effect can be seen when comparing
the top and bottom panels, although the effect is marginal for all but
the highest displayed accretion rates.

The left panels show accretion streams with an areal coverage of 10\%
and the left panels 1\%. As the areal coverage reduces, the effective
temperature of the accretion hot spot increases, resulting in an
increase in accretion flux, and a shift to bluer wavelengths of the
peak flux. This can be seen clearly by comparing the left and right
panels of Figure \ref{acc_dom}. As the rotation rate increases the
co-rotation radius also increases, as shown in Equation
\ref{inner_eq}, $R_{\rm inner}\propto \tau^{\frac{2}{3}}$. Therefore,
as Equations \ref{Lacc} and \ref{Tacc} show $L_{\rm acc}\propto 1-
\frac{R_*}{R_{\rm inner}}$ and $T_{\rm acc}\propto (L_{\rm
  acc}/A)^{\frac{1}{4}}$, the temperature of the accretion hot spot
increases as the rotational period increases. This is due to the
increase in potential energy lost by the mass accreted. This can be
seen by comparing the left and right panels of Figure
\ref{acc_dom}. Perhaps the most important, albeit qualitative, result
shown is Figure \ref{acc_dom} is an insight into the accretion rate at
which the accretion blackbody flux dominates over the photospheric
flux. Figure \ref{acc_dom} shows that as the accretion rate raises
above \logmdot = $-9$ for systems with 1 or 10\% areal coverage, the
accretion flux dominates the emergent SED at both
periods. Effectively, the spectroscopic features of the photosphere
are veiled by the additional continuum accretion flux. Therefore for
reasonable coverages (1--10\%) and rotational periods (0.5--5 days)
the photospheric flux is effectively veiled by accretion flux for
accretion rates \logmdot$>-9$. It is also clear from Figure
\ref{acc_dom} that the magnitude becomes brighter and the colour bluer
(in terms of optical photometry) with increasing accretion rates as
expected \citep{gullbring_1998}. The impact on the photometry of BDD
systems is discussed in the next section for individual stars and in
Section \ref{isochrones} for populations.

\subsubsection{Flaring and the Inner Edge}
\label{flare_inner_sec}

As shown in Figures \ref{flare_-12_183} and \ref{flare_-7_193} and
discussed in Section \ref{disc_flaring}, BDD systems in vertical
hydrostatic equilibrium have highly flared discs. As discussed in
\cite{walker_2004} this increased flaring (when compared to CTTS
stars) leads to occultation of the star at lower inclination angles
and, therefore, significant changes to the SED. Increases in the
inclination angle for these systems quickly lead to a significant
proportion of the stellar flux being intercepted and reprocessed by
the highly flared disc. This reprocessing will lead to a change in the
flux levels at the shorter, bluer, wavelengths as more stellar flux is
intercepted by the disc. It will also lead to significant changes in
the flux reaching the observer from the inner and outer regions of the
disc as the system approaches edge on. Also, as discussed in Section
\ref{disc_struct} the addition of dust sublimation leads to a change
in the shape of the inner edge, where the temperature is
sufficient. Figures \ref{no_sub_183_a}, \ref{sub_179_a} and
\ref{sub_181_a} show a range in inner edge shapes, from flat walls
through concave to convex curves, caused by the radial density profile
of the disc and the dependence of the sublimation temperature on
density. This change in shape, as noted for Herbig Ae stars by
\cite{tannirkulam_2007}, will lead to changes in the characteristics
of the SED, or derived IR excess, with inclination angle
\citep{tannirkulam_2007}, for close to face on viewing angles.

\begin{figure*}
  \centering
  \includegraphics[scale=0.6,angle=90]{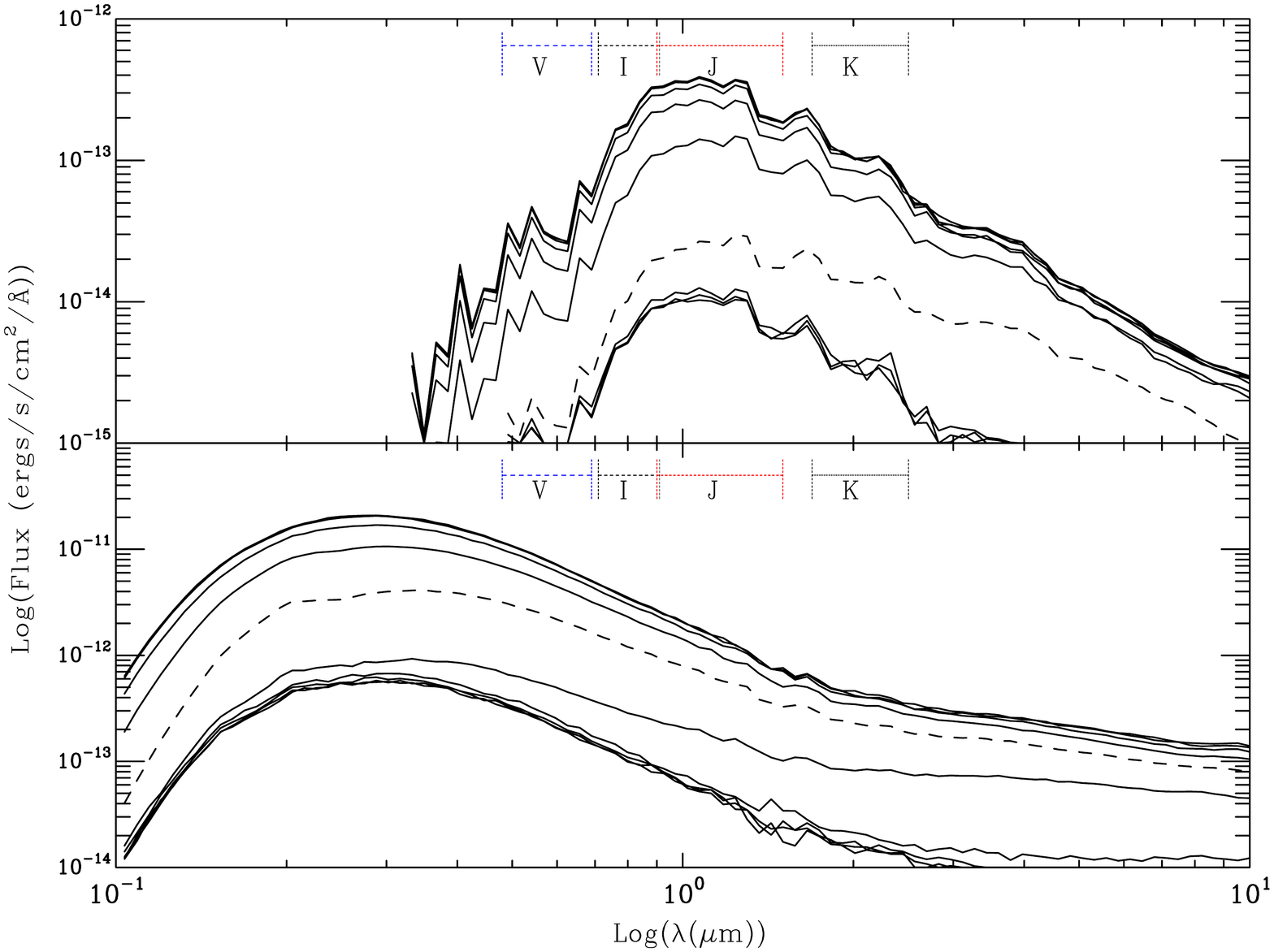}
  \caption{Figure showing both total SEDs of the systems shown in
    Figure \ref{flare_-12_183} and \ref{flare_-7_193} as the top and
    bottom panels respectively. The lines show the SEDs of all
    inclinations (see Table \ref{par_space_table}), with the
    obscuration angle shown as a dashed line, 71$^{\circ}$ and
    56$^{\circ}$ for \logmdot = $-$12 and $-$7
    respectively.\label{flare_inc}}
\end{figure*}

Figure \ref{flare_inc} shows the SEDs for the BDD system of Figures
\ref{flare_-12_183} and \ref{flare_-7_193} as the top and bottom
panels respectively. The lines show the flux over all the ten
inclinations (see Table \ref{par_space_table}), with the dashed line
showing the inclination at which a sharp fall in flux is seen. This
inclination will be the angle at which the star and inner disc becomes
obscured by the flared outer disc. These inclinations are 71$^{\circ}$
and 56$^{\circ}$ for \logmdot = $-$12 and $-$7 respectively. This
effectively means that for higher accretion rates a more significant
fraction of the stars, in a given population, may have flux below a
threshold detection limit.

\begin{figure*}
  \centering
  \includegraphics[scale=0.6,angle=90]{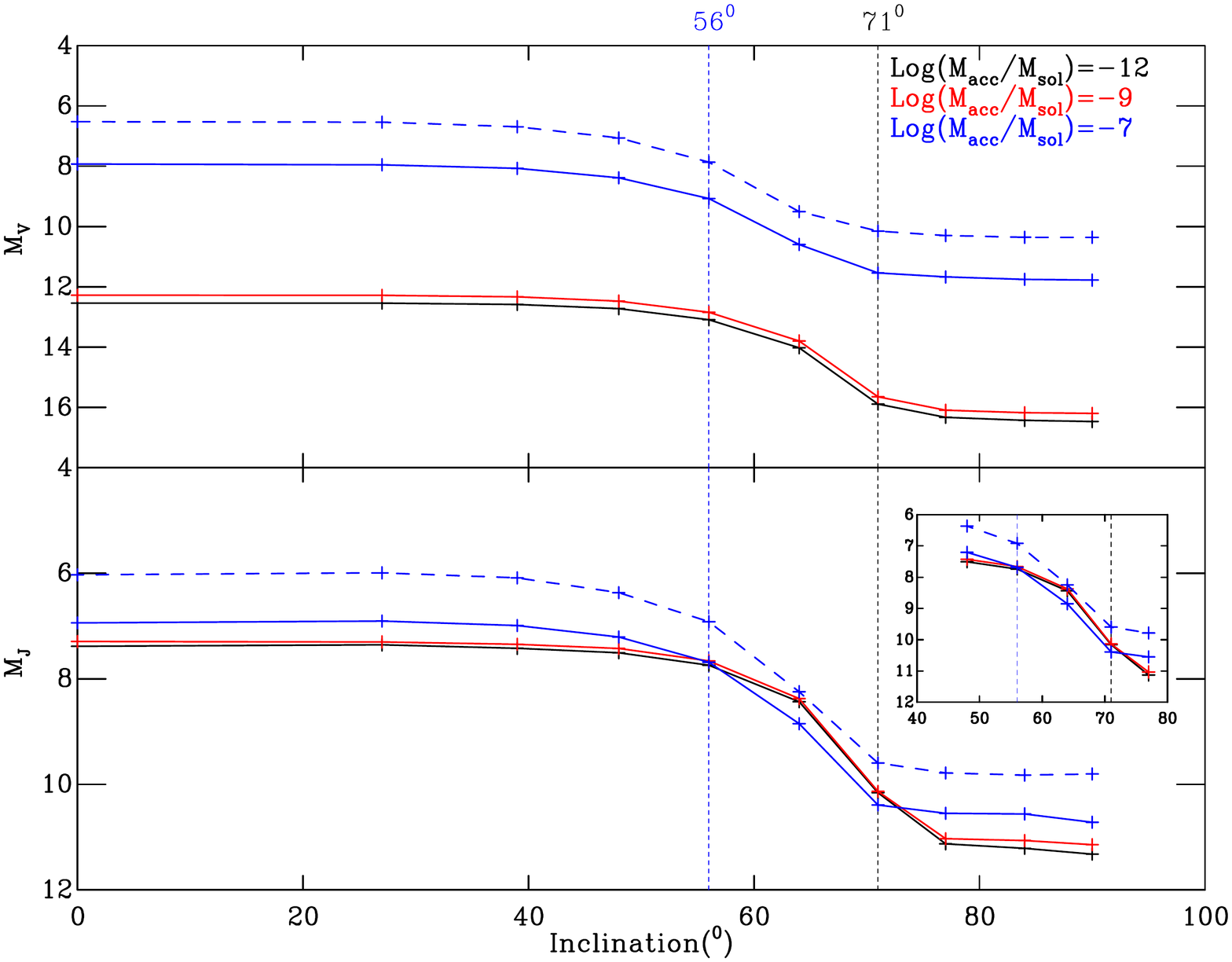}
  \caption{Figure showing $M_V$ and $M_J$ magnitudes as a function of
    inclination, shown as top, middle and lower panels
    respectively. The magnitudes of systems with $M_*=0.04M_{\odot}$,
    an age of 1 Myr, areal coverage of 10\% and rotation rate of 0.5
    days are shown as solid lines (and crosses). The systems with
    accretion rates of \logmdot = $-$12, $-$9 and $-$7, are shown in
    black, red and blue respectively. The panels also show the systems
    with the highest accretion rate but for a rotation rate of 5 days,
    as dashed lines (and crosses). The vertical dotted lines show the
    inclinations of 71$^{\circ}$ and 56$^{\circ}$. Finally, the inset
    panel simply shows a magnification of the region of interested for
    the $M_J$ magnitude. \label{flare_mag}}
\end{figure*}

The changes in flux with inclination will clearly lead to changes in
the observed magnitudes and colours with inclination. Figure
\ref{flare_mag} shows the $M_V$ and $M_J$ magnitudes in the top and
bottom panels respectively. The magnitudes are then marked as crosses
as a function of inclination. The systems shown in both panels have an
age of 1 Myr, mass of $0.04M_{\odot}$, rotation rates of 0.5 and an
areal coverage of 10\%. The systems with accretion rates of \logmdot =
$-$12, $-$9 and $-$7 are shown in black, red and blue,
respectively. The panels also show the system with the highest
accretion rate but with a rotation rate of 5 days as a dashed
line. The vertical dotted lines in the lowest panel simply illustrate
the inclinations of 71$^{\circ}$ and 56$^{\circ}$.

Accretion flux was shown to dominate the underlying, intrinsic,
photospheric SED for accretion rates of \logmdot = $-$9 (see Figure
\ref{acc_dom}). As one would expect this leads to significant changes
in the derived photometric magnitudes for bands blueward of a few
microns, where the accretion and photospheric flux dominate. This is
shown in the top panel of Figure \ref{flare_mag}. As the accretion
rate increases, the system moves to brighter magnitudes in $M_V$. The
move to brighter magnitudes becomes significant once the accretion
rate exceeds \logmdot = $-$9. Again moving from the fast to slow
rotation period increases the potential energy released and in this
case results in a brighter $M_V$ magnitude. As one moves to longer
wavelength photometric bands, as shown in the middle panel. The
increase in accretion rate affects these magnitudes, i.e. $M_J$ are
less significantly affected. Figure \ref{flare_mag} also shows the
change in magnitude caused by obscuration from the flared outer
disc. The occultation starts earlier for the higher accretion rates,
with the magnitudes in $M_V$ and $M_J$ moving fainter at earlier
inclinations for the higher accretion rates.

For our model grid the inner edge shape clearly changes with
increasing flux levels and therefore accretion rate. This, for
individual systems will lead to the reduced dependence of the IR
colours on inclination as the edge moves from a vertical to a convex
boundary as found by \cite{isella_2005}. Practically, for populations
of stars however disentangling this effect from the influences of
various other parameters (such as changing disc scaleheights and
accretion rates) on the IR colours is difficult. Essentially, matching
flux levels between two systems whilst maintaining a difference in
inner wall shape is impossible.

Thus, for individual systems changes in the accretion rate and disc
structure lead to significant changes in magnitudes and colours. The
data presented in this section are for an isolated mass and
age. However, the trends described are evident throughout our model
grid for any given subset, and as such are representative. The scatter
or changes in magnitude in the individual systems will change as a
function of the remaining variables, but the dominant input variables
affecting simulated photometry are accretion rate and inclination. For
populations of stars these changes in magnitude and colour act to
scatter or spread a BD locus in photometric space.

\subsection{Parameter derivation}
\label{isochrones}

Practically, most parameters for young pre-MS and BDs are derived
from surveys of populations, usually open clusters, using broadband
photometry and subsequently constructed colour-magnitude and
colour-colour diagrams (CMDs and CoCoDs respectively, hereafter).  In
this section we explore the consequences of our model grid on the
derivation of the primary parameters of age, mass and disc fractions,
from populations. This in turn leads to highlighting selection effects
with, for instance, the mass to accretion rate relation.

To delineate the effects of the accretion rate and circumstellar discs
we have subdivided the grid into two groups (as discussed in Sections
\ref{disc_struct}), those with accretion rates typical for higher mass
CTTS objects, defined as \logmdot = $-$12 (negligible) to \logmdot =
$-$9 and those with elevated accretion rates, where \logmdot
$>-$9. For several of the plots in this section the magnitude and
colours for the $M_*=0.01M_{\odot}$ systems at high inclinations
become extremely faint and red, and we omit them from the figures in
order to conserve an appropriate scaling for the majority of the
models. Additionally, stars at the highest inclinations, i.e edge on
disc systems, are often omitted from the figures due to their
extremely faint magnitudes, meaning they would not be practically
observable.

As discussed in Section \ref{disc_struct} the inner edge location is
correlated with inner edge temperature (albeit differently for the
typical and extreme accretors). As noted by \cite{meyer_1997} this
could lead to a correlation of IR excess with inner edge position. As
expected however, scatter caused by variations in inclination and disc
structure act to remove this correlation for populations.  The
correlation for an individual set of a systems, i.e. all variables
fixed except rotation rate, between IR colour and rotation rate is
preserved. However, as shown in Section \ref{disc_struct} the changes
caused by accretion rate and inclination angle are the most
significant, and act to remove any correlation between rotation rate
and IR colours for populations. In fact, for our grid the simulated
photometry shows no significant correlation between rotation rate and
IR colours. The work of \cite{meyer_1997} studied flat accreting disc,
therefore, as we have increased variation in disc structure by
applying vertical hydrostatic equilibrium it would be interesting to
see if any correlation appears for analytically defined disc
structures. We have run a set of parallel models using analytical
disc structures and will publish the results in a subsequent
publication. Overall, as found in \cite{walker_2004}, the dominant
scattering effect for BD disc systems in an optical CMD appears to be
caused by accretion rate and inclination. Therefore for derivation of
parameters we explore the scatter in BD loci as a function of
accretion rate and inclination.

\subsubsection{Mass and Age Derivation}
\label{mass_age}

For the derivation of ages optical CMDs, in particular in \textit{V,
  V$-$I}, are most often used, and indeed most suitable. Whereas, IR
CMDs, such as a \textit{J, J$-$K} CMD, are most suitable for mass
derivation \cite[see references and discussions in][ and Section
\ref{derived}]{mayne_2007,mayne_2008}.

The use of pre-MS and BD isochrones for the derivation of single star
parameters is at present not proven to be reliable \citep[see
discussion in][]{mayne_2008}. Practically, therefore, median ages are
derived from populations. Subsequently, derived masses are still
unreliable but at least based on a consistent age. This problem is
being addressed by Bell et al (in prep), where $K$ band photometry and
known eclipsing binaries are being used to refine pre-MS
isochrones. In this section we plot the data for our 1 Myr systems
only and explore the resulting scatters caused by the disc presence
and accretion luminosity.

\begin{figure*}
  \centering
  \includegraphics[scale=0.6,angle=90]{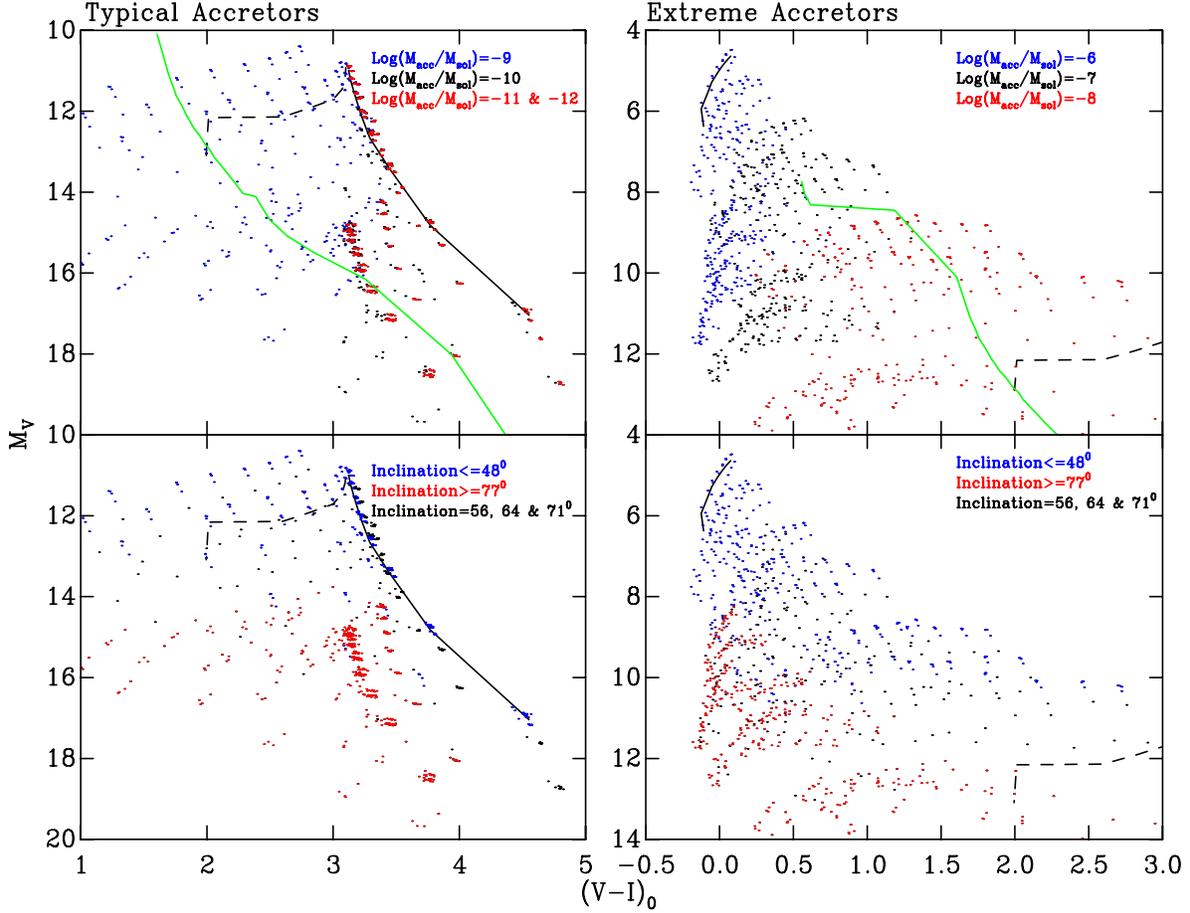}
  \caption{Figure showing CMDs in $M_V$, $(V-I)_0$ for typical
    accretors (\logmdot $\leq$ $-$9) in the left panels, and extreme
    accretors (\logmdot $>-9$) in the right panels. The black dashed
    and solid lines correspond to naked BD isochrones (from our grid)
    with \logmdot\ $-$9 and $-$12 for the left panels (typical
    accretors) and -9 and -6 in the right panels (extreme
    accretors). The rotation rates and areal coverages are set at 5
    days and 10\% respectively (changing these has little effect). The
    top panels also include as a solid green line the 1 Myr pre-MS
    isochrone of \citet{siess_2000} adjusted to a distance modulus of
    7 and an extinction of $A_V=2$, simulating a reddened background
    population. The top panels then separate the systems by accretion
    rate with \logmdot = $-$9, $-$10 and $-$11 \& $-$12 in the top left panel,
    and \logmdot = $-$6, $-$7 and $-$8 in the top right panel, shown blue,
    black and red dots respectively for both cases. The lower panels
    then split the systems into groups by inclination, with $\theta
    \leq$ 48$^{\circ}$, $\theta \geq$ 77$^{\circ}$ and $\theta=$ 56,
    64 \& 71$^{\circ}$, plotted as blue, red and black dots
    respectively.\label{age_mass_VVI}}
\end{figure*}

\begin{figure*}
  \centering
  \includegraphics[scale=0.6,angle=90]{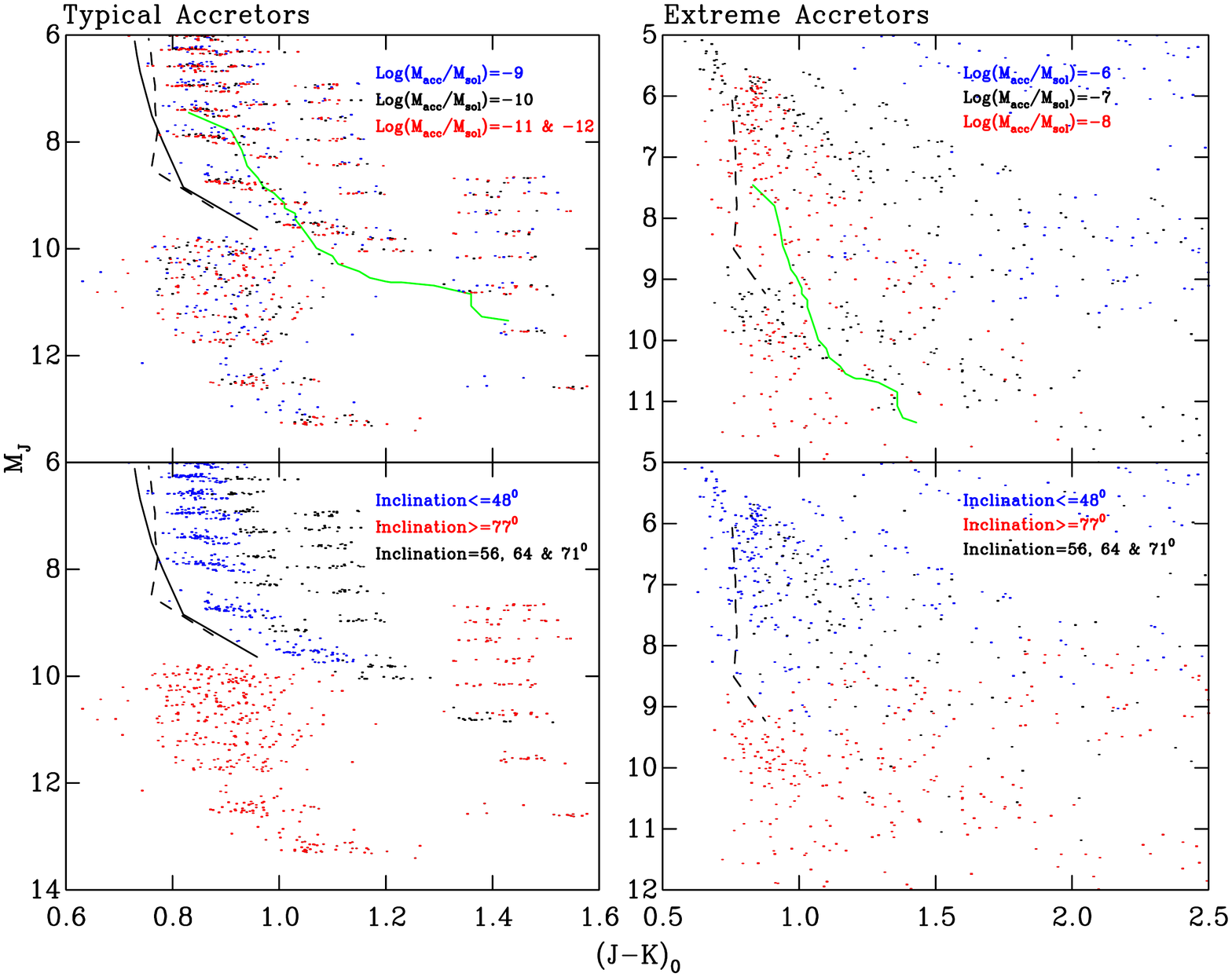}
  \caption{Figure showing the same data in the same format and with
    the same symbol meanings as Figure \ref{age_mass_VVI}, but for
    $M_J$ and $(J-K)_0$ CMDs. However, for this Figure the \logmdot =
    $-$6 isochrone is not shown (lies significantly blueward of locus
    of data), and the pre-MS isochrone of \citet{siess_2000}, has been
    smoothed.\label{age_mass_JJK}}
\end{figure*}

Figures \ref{age_mass_VVI} and \ref{age_mass_JJK} shows CMDs in $M_V$
\& $(V-I)_0$ and $M_J$ \& $(J-K)_0$ respectively. The left panels of
both figures shows stars classed as typical accretors with accretion
rates of \logmdot = $-$9, $-$10 and $-$11 \& $-$12, shown as blue,
black and red dots respectively in the top panels. The right panels of
Figures \ref{age_mass_VVI} and \ref{age_mass_JJK} show systems with
extreme accretion rates, with \logmdot = $-$6, $-$7 and $-$8 shown as
blue, black and red dots respectively, in the top right panel. The
bottom panels then show the systems separated into groups by
inclination. These groups are $\theta \leq 48^{\circ}$ as blue dots
(classed as face-on systems), $\theta> 56$ \& $64^{\circ}$ (classed as
the expected systems, as the expectation value of
$cos(\theta)=60^{\circ}$) as black dots and $\theta \geq 71^{\circ}$
(classed as edge-on systems) as red dots (for typical, bottom left,
and extreme, bottom right, accretion rates). The left panels then show
naked stars isochrones (created from our grid of simulated photometry)
for 1 Myrs at accretion rates of \logmdot = $-$12 and $-$9, as solid
and dashed lines respectively (with an areal coverage of 10\% and
rotation rates of 5 days). Whereas the left hand panels show naked
star isochrones of \logmdot = $-$6 and $-$9 as solid and dashed black
lines respectively (with an areal coverage of 10\% and rotation rates
of 5 days). The solid green line, in the top panels, shows the 1 Myr
isochrone of \cite{siess_2000} adjusted to a distance of 250 pc and an
extinction of $A_V=2$ mag, simulating a background population of CTTS
stars.

As can be seen in Figures \ref{age_mass_VVI} and \ref{age_mass_JJK}
current accretion and disc presence in a star and disc systems creates
a scatter in our simulated photometry indicative of a large isochronal
age or mass spread. Indeed, for many BDD systems even at nominal
accretion rates of \logmdot = $-$11 or $-$12, for our simulations, the
colours of these stars move significantly blueward of the expected BD
locus in a $M_V$, $(V-I)_0$ CMD (and redward in a $M_J$, $(J-K)_0$
CMD). As, in the case of typical accretors, the input variables are in
the range expected for a BD population, one could reasonably expect
observed true BD loci to show a similar scatter. It is clear that even
a wide photometric selection would not even include all of the
negligibly accreting systems at expected inclinations. Furthermore,
the derivation of masses and ages for these objects will be
difficult. Additionally, for the higher accretion rates of \logmdot =
$-$9 or $-$10 the movement of the star within the CMD will effectively
shift the star into the contamination region expected for background
CTTS or MS stars at a $(V-I)_0$ of $\leq$1.5 and $M_V$ 12--10, and as
such the star would not be included in a photometrically selected BD
sample. The solid green line, in the top panels of Figures
\ref{age_mass_VVI} and \ref{age_mass_JJK}, showing the 1 Myr isochrone
of \cite{siess_2000} at a distance of 250 pc and extinction of $A_V=2$
mags, shows that the BDD systems with higher accretion rates could
easily be confused for a background CTTS or MS population. Indeed
misclassification of a BDD system as a CTTS system has already been
revealed in \cite{white_2003}. For the extreme accretors, as shown in
the right panels of Figures \ref{age_mass_VVI} and \ref{age_mass_JJK},
there is little chance of these objects being classed,
photometrically, as BD candidates or being assigned the correct
mass. Therefore, if one attempted to locate a population of BDD
systems with elevated or extreme accretion rates, target selection
would have to be placed at much brighter magnitudes, and bluer for
optical or redder for IR, colours.

This scatter for both typical and extreme accretors is a strong
function of inclination, where, as the inclination is increased the
objects are pushed lower in the CMD. Indeed, for the edge on cases
some objects have magnitudes fainter than those shown (for instance
$M_V\approx$\,20). This is expected as the star becomes obscured by the
flared disc, interestingly for typical systems the bottom right panels
show that this occurs for inclinations above around 71$^{\circ}$ (as
found in Figure \ref{flare_inc}) in most cases.  However, even for the
lower inclination angles some objects have very faint magnitudes, this
is due to the disc flaring leading to earlier obscuration of the star
as discussed in Section \ref{disc_struct}. Crucially, the top panels
show that the scatter from the isochrone is generally correlated
with accretion rate. Effectively, as the accretion rate increases the
BDD system moves farther away from the isochrone and is therefore less
likely to be classified as a BDD system and included in any target
samples of such objects. Overall, the dominant scattering effect for
BDD systems in an optical CMD appears to be caused by accretion rate
and inclination, and therefore obscuration effects of the disc on the
star.

We have presented CMDs constructed using $M_V$, $(V-I)_0$ and $M_J$,
$(J-K)_0$. The scatter and correlations of scatter with accretion rate
and inclination found within these CMDs are however, representative of
CMDs constructed using optical or near-IR colours and magnitudes.

If one adopts the range of input parameters we have used (see Section
\ref{par_space} for justification), our simulated photometry shows,
qualitatively, that a coeval 1 Myr population of accreting BDs
and BDD systems, with typical accretion rates and range of
inclinations, will exhibit a significant scatter in apparent
isochronal age. Furthermore, objects with typical (and extreme)
accretion rates are scattered sufficiently in CMD space to prohibit
their identification as BDs. Indeed, these objects would not be
included in a photometrically selected sample of BDs, and as such are
unlikely to be assigned the correct masses or ages. The scatter from
the naked 1 Myr systems, generally, increases with increasing
accretion rate. It is important to note at this point that these
conclusions are qualitative, and obviously based on our
assumptions. However, the repercussions for isochronal age derivation
and sample selection in the BD regime could be profound. Indeed this
study has only included the effects of current or ongoing accretion
from a disc, it has neglected any effects of accretion, both past and
present, on the evolution of the central star. This past accretion
could also act to reduce the stars radius, accelerating contraction
\citep{tout_1999,siess_1999} and introducing additional scatter in a
coeval population proportional to the range in accretion rates
\citep[see][for full discussion]{mayne_2008}. In addition we have
shown that the typical accretion rates may scatter BDD systems into
the region of a CMD occupied by the CTTS or background MS locus.

The fact that scatter in the CMD increases with increasing accretion
rate casts doubt on the veracity of the mass to accretion rate
relationship. For our model grid we have not assumed any such
relation, therefore, as our data would also show a similar relation it
suggests that the observed result may be caused by intrinsic
scattering. The relation $\dot{M}\propto M_*^{2}$ suggests that their
is a dearth of lower mass stars accreting at higher rates. We have
shown that for accretion rates in the range \logmdot = $-$12 to $-$9
the BDD systems with higher accretion rates would be preferentially
missed using photometric or isochronal selection. Furthermore, for the
extreme accretors the BDD systems are scattered far from the
non-accreting naked BD locus. This effectively means that applying
standard photometric selection and considering possible dynamic
magnitude ranges (to define saturation and magnitude limits), the
systems with extreme accretion rates would not be classified as BDD
systems.

\subsubsection{Disc Fractions}

Disc fractions have been derived using infrared excesses previously in
\textit{JHK} \citep[e.g][]{lada_1995,carpenter_1997}, however recent
works pre-dominantly use \textit{Spitzer} IRAC magnitudes
\citep[e.g][]{luhman_2005b,luhman_2008,gutermuth_2008,monin_2010}. Furthermore,
MIPS magnitudes are used to identify so-called debris discs, where IR
excesses are not apparent at shorter wavelengths
\citep[e.g][]{currie_2008,bryden_2009}. Finally, disc fractions have
also been derived using the $\alpha$ criteria, where
$\alpha=\frac{dlog\lambda F\lambda}{dlogF\lambda}$ between two
limiting wavelengths, originally used to distinguish amongst Class I,
II or III sources, but now used to detect disc presence
\citep{lada_2006,kennedy_2009}. An $\alpha>-2$ is used as a selection
criterion for disc presence for TTS stars. We have constructed the
$\alpha$ values for our model grid by adopting the limiting
wavelengths of \cite{kennedy_2009}, namely 3.6 to 8.0$\mu$m.

\begin{figure*}
  \centering
  \includegraphics[scale=0.6,angle=90]{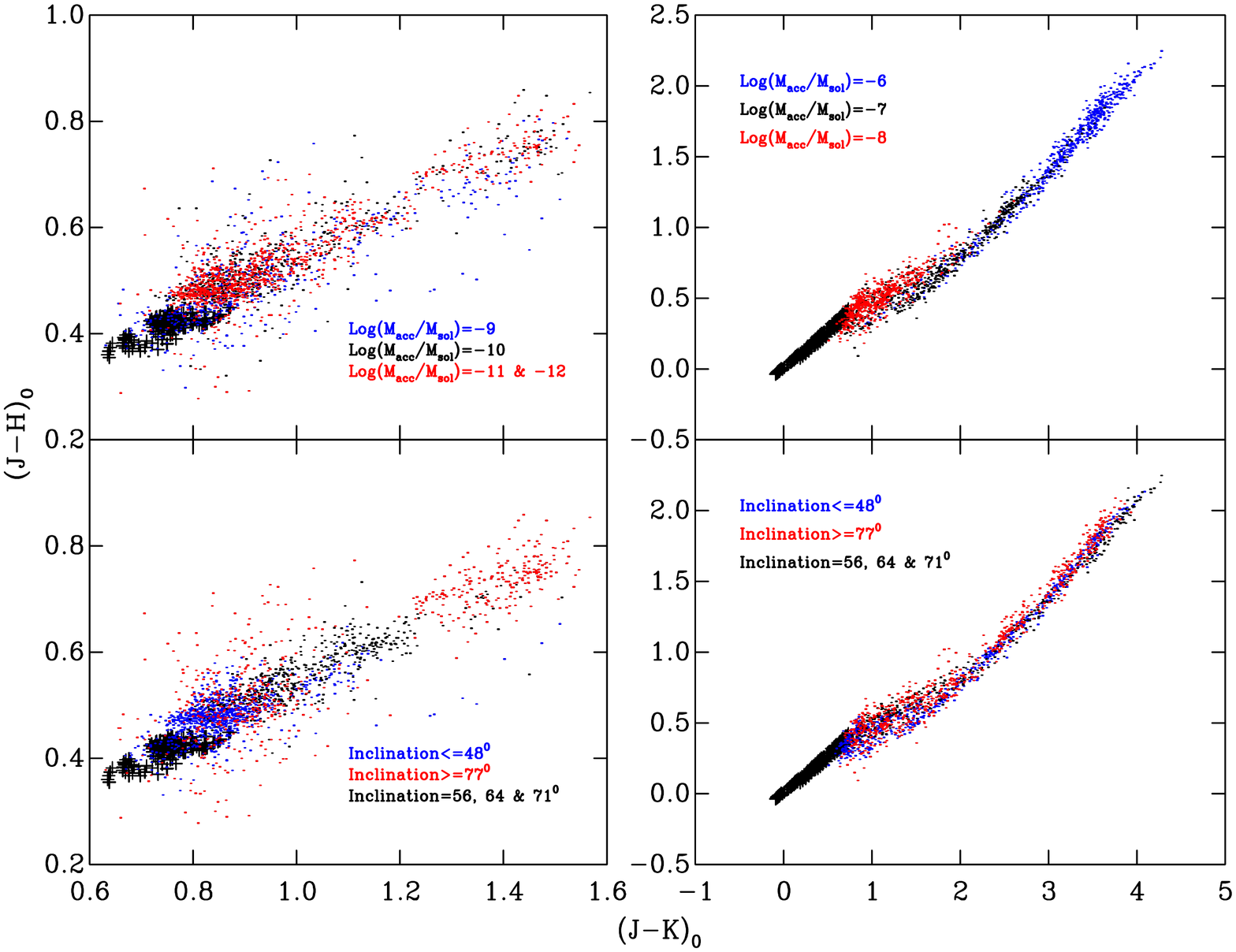}
  \caption{Figure showing $(J-K)_0$, $(J-H)_0$ CoCoDs for typical
    (left panel) and extreme accreting (right panels) systems. The top
    panels then separate the accretion rates with \logmdot = $-$9,
    $-$10 and $-$11 \& $-$12, and \logmdot = $-$6, $-$7 \& $-$8 shown
    as blue, black and red dots respectively for the typical (top left
    panel) and extreme (top right panel) accretors. The bottom panels
    then show the systems with the inclinations $\theta \leq
    48^{\circ}$ as blue dots, $\theta> 56$ \& $64^{\circ}$ as black
    dots and $\theta \geq 71^{\circ}$ as red
    dots.\label{disc_frac_JHJK}}
\end{figure*}

Figure \ref{disc_frac_JHJK} shows the photometry for all models in our
grid in a $(J-K)_0$, $(J-H)_0$ CoCoDs. In all panels, all of the naked
systems are shown as black crosses. The left panels show those systems
with typical accretion rates and the right panels the extreme
accretors. The top panels of Figure \ref{disc_frac_JHJK} then separate
the systems by accretion rate with \logmdot = $-$9, $-$10 and $-$11 \&
$-$12, and $-$6, $-$7 and $-$8, shown as blue, black and red dots
respectively for the typical (top left panel) and extreme (top right
panel) accretors.  The bottom panels then shows the defined groups of
inclination angles, with $\theta \leq 48^{\circ}$ as blue dots,
$\theta> 56$ \& $64^{\circ}$ as black dots and $\theta \geq
71^{\circ}$ as red dots.

Figure \ref{disc_frac_JHJK} shows that there is, for both typical and
extreme accretors, an overlap between the naked and BDD systems. This
suggests that an empirically placed cut in a CoCoD of this type (in
the absence of complications from variable reddening) could
mis-identify some candidates. However, as disc fractions defined by
placing a colour-colour cut are viewed as lower limits this effect may
be small. The photometry from our grid shows no correlation in a CoCoD
of this type with rotation rate. In the case of the extreme accretors
(right panels) there is a strong correlation of accretion rate and
scatter from the naked stars. Additionally, the extreme accretors lie
significantly removed from the naked stars locus. There is a
possibility that some of these objects may be lost due to saturation
and limiting magnitude effects. Again, as with mass and age
derivation, the present figures are representative of similar figures
using alternative, passbands within the instrument filter sets.

\begin{figure*}
  \centering
  \includegraphics[scale=0.6,angle=90]{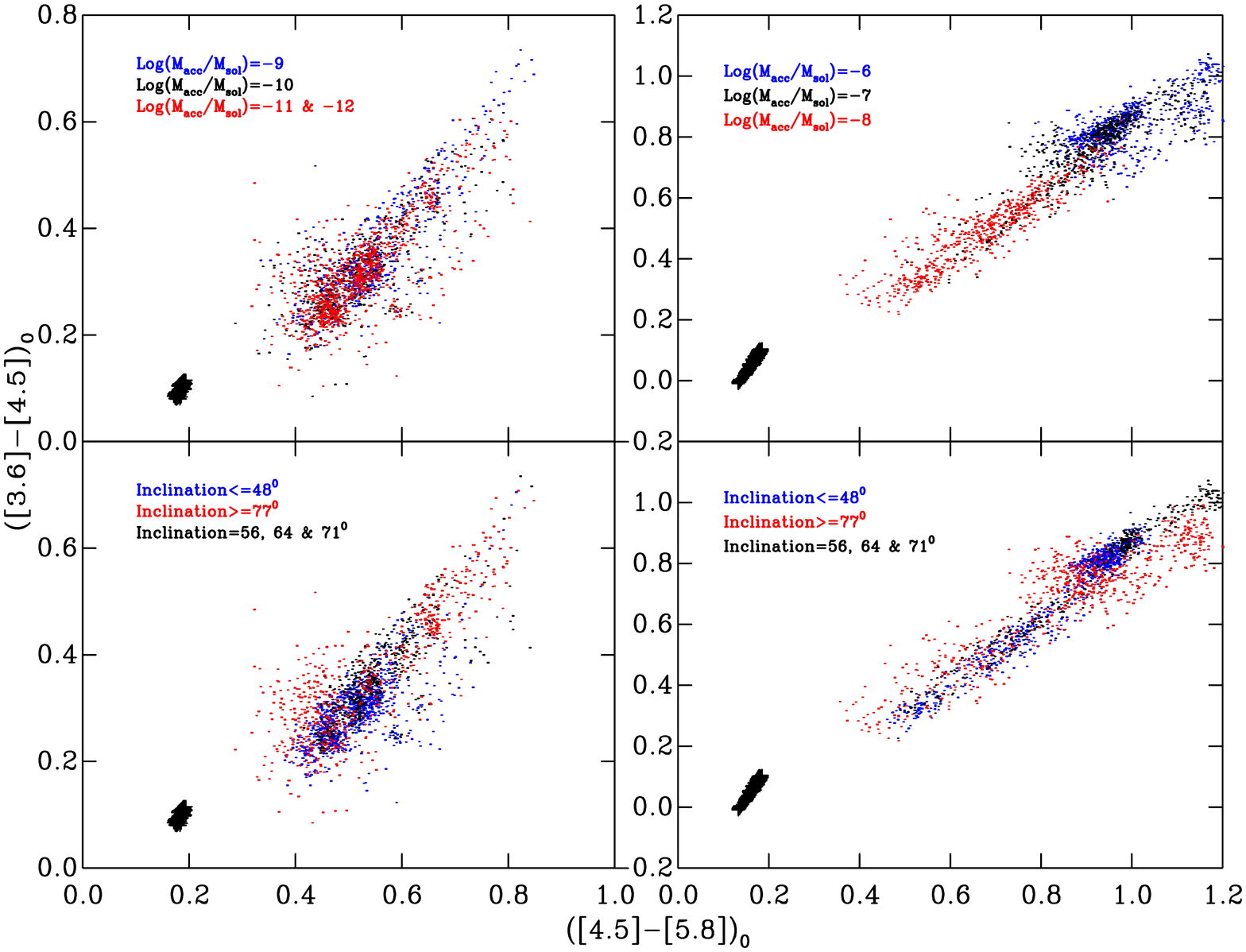}
  \caption{Figure showing $([3.6]-[4.5])_0$, $([4.5]-[5.8])_0$ CoCoDs
    of both typical (left panels) and extreme (right panels)
    accretors. The panels separate the different accretion rates and
    inclinations as in Figure
    \ref{disc_frac_JHJK}.\label{disc_frac_irac}}
\end{figure*}

\begin{figure*}
  \centering
  \includegraphics[scale=0.6,angle=90]{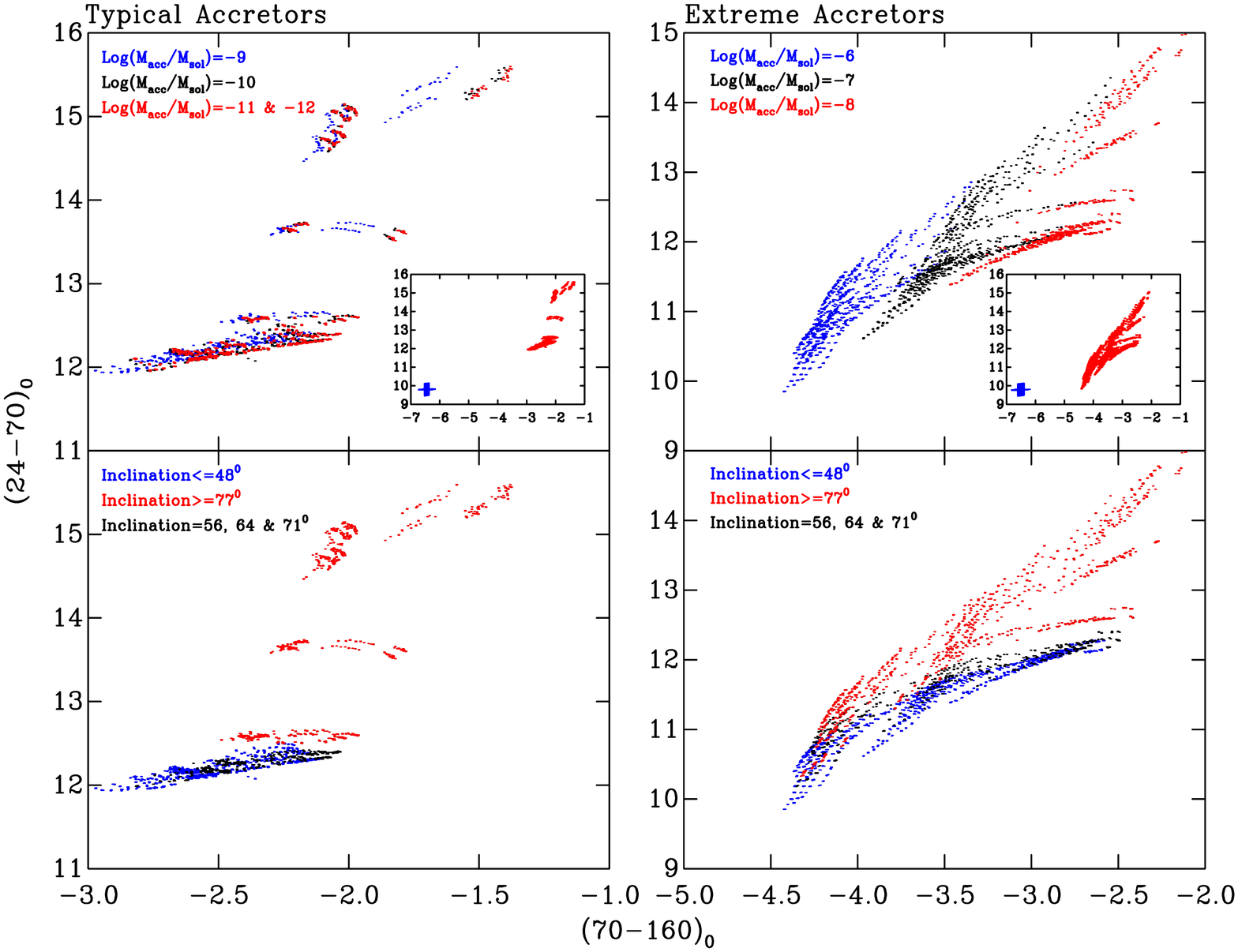}
  \caption{Figure showing $(24-70)_0$, $(70-160)_0$ CoCoDs in the same
    format as Figure \ref{disc_frac_JHJK}. The smaller inset panels in
    the top panels show the separation between the naked (blue
    crosses) and BDD systems (red dots).\label{disc_frac_mips}}
\end{figure*}

CoCoDs constructed using the longer wavelength bands of the IRAC and
MIPS cameras are most commonly used for disc fraction
calculation. Selection based on these types of data are well accepted
as indicators of disc presence. Figures \ref{disc_frac_irac} and
\ref{disc_frac_mips} show example CoCoDs for IRAC and MIPs magnitudes
respectively.

Figures \ref{disc_frac_irac} and \ref{disc_frac_mips} show the same
data in the same format with the same symbol meanings as Figure
\ref{disc_frac_JHJK}, except the colour indices. Figure
\ref{disc_frac_irac} shows $([3.6]-[4.5])_0$, $([4.5]-[5.8])_0$ CoCoDs
and Figure \ref{disc_frac_mips} shows a $(24-70)_0$, $(70-160)_0$
CoCoDs. Figure \ref{disc_frac_mips} also shows, as insets within the
top panels, a larger scale figure to include the naked systems.

Figure \ref{disc_frac_irac} shows that, as expected, the IRAC CoCoDs
clearly separate the naked and BDD systems. Once again, as found in
Figure \ref{disc_frac_JHJK}, for the extreme accretors the separation
of the BDD systems from the naked stars is a weak function of
accretion rate. There is also a weak correlation with inclination,
with the face-on and expected systems closer to the naked
stars. Figure \ref{disc_frac_irac} is representative of CoCoDs
constructed using other IRAC photometric channels. As we include
longer wavelength bands in the IRAC CoCoDs the separation between the
disc and naked loci increases. As the wavelength gets longer the flux
originates from regions of the disc at lower temperature and therefore
greater radial distance from the star. As photometric emission is
minimal past around 3\,$\mu$m, systems with discs will have
significantly different SEDs from naked systems at these wavelengths.

Figure \ref{disc_frac_mips} shows that as we increase the wavelength
even further, into the range of the MIPS photometric bands, the
separation between naked and BDD systems increases still further. The
inset panels show that the separation between the naked and BDD
systems is larger than in Figure \ref{disc_frac_irac}. Additionally,
there are clear correlations of MIPS positions with accretion rate and
inclination. For the extreme accretors the systems are clearly
delineated by accretion rate and inclination. This is as the emission
is coming from greater radial positions within the disc where the
structure of the disc is a finer function of the input
variables.

Therefore, for models within our grid, disc fractions can be easily
derived using IRAC and MIPS data. Whereas, $JHK$ data used to derive a
disc fraction will lead to a probable underestimate of the disc
fraction. There also appears a correlation, especially for the extreme
accretors, with position in the CoCoDs and accretion rate (and perhaps
inclination).

\subsubsection{Observational cuts}
\label{cuts}

Recent derivations of disc fractions usually use colour-colour
selection in the IRAC photometric bands. As we have shown within our
model grid the separation between the BDD and naked systems is
clear. Therefore, we can examine the success of some recent
observationally placed cuts when applied to our model
grid. Additionally, disc fractions have been derived using the
$\alpha$ value \citep{lada_2006}, essentially a slope of the SED
between two wavelengths (at wavelengths longer than the stellar flux
peak). As these $\alpha$ values are usually derived at wavelengths
across the IRAC bands one would expect the resulting disc fractions to
be reliable.

\begin{figure*}
  \centering
  \includegraphics[scale=0.65,angle=90]{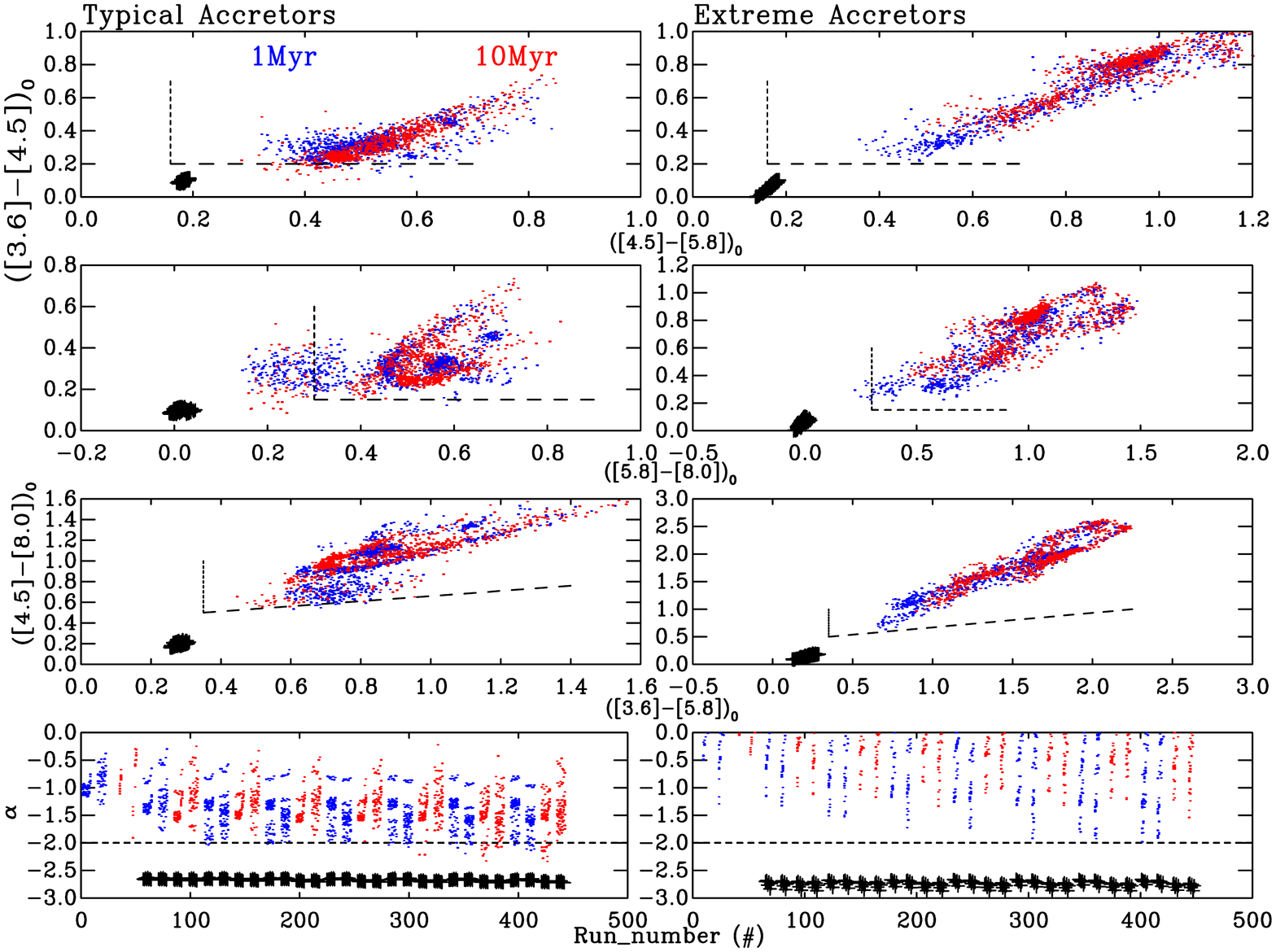}
  \caption{Figure showing both from top to bottom, the disc candidate
    selection of \citet{luhman_2005b}, \citet{luhman_2008},
    \citet{monin_2010} \citep[using the criteria of][]{gutermuth_2008}
    and \citet{kennedy_2009}. The photometry or $\alpha$ values for
    all models are shown, with naked stars as black crosses, 1 Myr BDD
    systems as blue dots and 10 Myr BDD systems as red dots. The
    dashed horizontal and vertical lines are disc selections from the
    studies in question. The left panels are the typical accretors and
    the right panels the extreme accretors. The top three panels are
    IRAC CoCoDs and the bottom panel plots the $\alpha$ value against
    run number (an arbitrary number).\label{disc_cuts}}
\end{figure*}

Figure \ref{disc_cuts} presents all the data for our model grid
separated by age, with 1 and 10 Myr BDD systems shown as blue and red
dots respectively. The naked systems are shown as black crosses. The
colour combinations used to construct the CoCoDs featured are from
several recent publications where disc fractions have been
derived. The selection criteria from theses studies are marked as
dashed lines. The left panels show the typical accretors and the right
panels the extreme accretors.

The observational cuts applied, shown as dashed lines, are from
\cite{luhman_2005b} a study of IC348, \cite{luhman_2008} a study of
$\sigma$ Orionis and \cite{monin_2010} a study of the Taurus region
\citep[using the criteria of][]{gutermuth_2008}. In these cases the
effects of extinction are either negligible in the plotted colours,
with values of $E([3.6]-[4.5])<0.04$ and $E([4.5]-[5.8])<0.02$ for
IC348 and $A_V \leq$4 mag \citep[which will be negligible in the IRAC
CoCoD,][]{allen_2004}, or the cuts have been placed in intrinsic
colour space as for $\sigma$ Orionis. The cuts from
\cite{luhman_2005b}, \cite{luhman_2008} and \cite{monin_2010} appear
as the top, top-middle and bottom-middle panels. The lower panel shows
a recent BDD candidate selection using the $\alpha$ value from
\citep{kennedy_2009}.

Figure \ref{disc_cuts} shows that the recent observational cuts used
to identify disc candidates would be, in the main, reliable for our
model grid. Simulated photometry would be correctly identified using
these cuts. It is important to note that our conclusions so far have
been drawn from differential photometric arguments, in this case we
are using intrinsic colours and these values are extremely sensitive
to changes in zero point and photometric calibration. The top panel of
Figure \ref{disc_cuts} shows that other than a few typically accreting
systems the observational cut of \cite{luhman_2005b} would select all
of the BDD systems in our grid. The second panel down shows that the
cut of \cite{luhman_2008} would miss some typical and a very small
number of extreme accreting BDD systems (almost all are edge-on
systems). The third panel down shows only very few BDD systems will be
missed by the selection of \cite{gutermuth_2008} which was applied to
Taurus by \cite{monin_2010}. Finally, the selection for CTTS candidates
of $\alpha>-$2 \citep{kennedy_2009} appears reasonably applicable to
our BDD systems. As can be seen in the bottom panel of Figure
\ref{disc_cuts} almost all the typical, and all of the extreme
accreting BDD systems would be successfully identified using this
criterion. Again the missed typically accreting systems are edge-on
systems. Given, that extreme reddening may move naked stars into the
BDD region selected, in general, disc fractions are usually quoted as
lower limits. Therefore, a small number of miss-identified BDD system
is a negligible effect. This allows us to conclude, that ubiquitously
used BDD selections within IRAC CoCoDs are reliable when applied to
our model grid. Additionally, the $\alpha$ value would be a reliable
disc indicator for our model grid.

Using our model grid we can define the optimal cuts for selecting BDD
systems. These are shown in Table \ref{cuts_table}, where the defined
cuts are chosen to minimise contamination from naked stars.

\begin{table}
\begin{tabular}{|l|l|l|}
\hline
Colour index&Disc selection&Number missed (\# \& \%)\\
\hline
$\alpha(3.6-8.0)_0$&$> -2.20$&26/4480=0.6\%\\
$([3.5]-[4.5])_0$&$> +0.21$&201/4480=4.5\%\\
$([3.5]-[5.8])_0$&$> +0.50$&12/4480=0.3\%\\
$([4.5]-[5.8])_0$&$> +0.32$&9/4480=0.2\%\\
$([4.5]-[8.0])_0$&$> +0.45$&7/4480=0.2\%\\
$([5.8]-[8.0])_0$&$> +0.13$&13/4480=0.3\%\\
$([8.0]-24)_0$&$> -16.2$&0/4480=0.0\%\\
$(24-70)_0$&$> +9.90$&2/4480=0.04\%\\
$(70-160)_0$&$> -6.20$&0/4480=0.0\%\\
\hline
\end{tabular}
\caption{List of all the cuts which when applied to our simulated
  dataset provide the best disc candidate selection with minimised
  contamination. \label{cuts_table}}
\end{table}

\section{Conclusions}
\label{conclusions}

We have constructed a model grid of SEDs, and subsequently photometric
magnitudes and colours, for actively accreting BDs with or without an
associated accretion disc. We have modelled the photospheric flux from
these BDs by adopting (and interpolating) the interior `DUSTY00'
models of \cite{chabrier_2000} combined with the `AMES-Dusty',
atmospheric models of \cite{chabrier_2000}. We have then assumed that
accretion occurs from an inner edge of a magnetically truncated
accretion disc (truncated at the co-rotation radius). The accretion
flux is calculated using a simple blackbody emission, given the
derivation of a characteristic spot effective temperature. SEDs were
then produced for both naked BDs and BDD systems. For the BDD systems
we have modelled the disc using the {\sc torus} radiative transfer
code. The {\sc torus} code implements the \cite{lucy_1999} radiative
equilibrium algorithm, incorporates dust sublimation and includes a
treatment of vertical hydrostatic equilibrium (see Section \ref{model}
for a discussion of the code). In order to produce a grid of simulated
systems we have varied several input parameters namely: stellar mass,
stellar age, stellar rotation rate, accretion rate, the areal coverage
of the accretion stream and the system inclination (the disc mass was
fixed). The ranges of these variables were selected to represent and
bound typical BD systems, justification is provided using evidence
from observational studies in Section \ref{par_space} and a final list
of the values of these variables can be found in Table
\ref{par_space_table}.

Accepting our assumptions, parameter ranges and radiative transfer
code our resulting simulated dataset has allowed us to qualitatively
explore the effects of \emph{active} (current not past) accretion on
disc structure. Furthermore through the simulation of observations we
have explored the effects of accretion, and disc presence, on both the
SEDs, and photometric colours and magnitudes of these systems.

As discussed in Section \ref{disc_struct} vertical hydrostatic
equilibrium, when applied to BDs, leads to increased flaring, when
compared to CTTS. This has previously been explored by
\cite{walker_2004}. However, in our study we have included a simple
treatment of accretion. This leads to increased flaring as more flux
reaches the outer disc, and subsequently lower inclinations angles at
which the central star is obscured for BDD systems with higher
accretion rates. Furthermore, the addition of dust sublimation has
shown that for BDD systems the inner disc location, temperature and
vertical size and shape also varies with accretion rate. The inner edge
position is correlated with temperature for the lower accreting models
as suggested by \cite{meyer_1997}. For the systems with higher
accretion rates the inner edge temperature is weakly correlated with
temperature, mainly due to the radial fall in density and therefore
dust sublimation temperature. The inner disc edge, initially
prescribed as a vertical wall, then becomes concave and finally convex
as dust sublimation is increased (with increasing flux from higher
rates of accretion).

Subsequently the SEDs of BDD systems with typical accretion rates and
associated discs change significantly from the assumed underlying
photospheric model flux, and therefore become difficult to
classify. In Section \ref{observables} we have shown that the BD
photosphere becomes veiled by the accretion flux for rates of
\logmdot$>-$9. The outer disc flaring observed in the BDD systems was
shown to cause occultation and a subsequent, sharp, fall in flux at an
inclination which decreases for systems with higher accretion
rates. We have also shown that for extreme accretion rates the inner
wall increases in size and becomes convex in shape.

Subsequent derivation of photometric magnitudes has allowed us to
demonstrate that, as expected, increased accretion without disc
presence, moves our naked systems to bluer and brighter magnitudes.
Once a disc is added the increase in accretion flux interacts with the
disc and does not necessarily lead to a simple motion toward brighter
magnitudes and bluer colours. The increased flaring and obscuration
present in BDD systems, over CTTS, leads to rapid falls in magnitude
with inclination as an accretion (or flaring) dependent inclination.
Furthermore, the disc inner edge leads to a shift redwards with
increasing accretion rate as more flux is intercepted by the inner
edge and the inner edge becomes convex and `puffed up'.

In practice most parameters for BDD systems are derived for
populations. We have shown, in Section \ref{isochrones} that
derivation of an \emph{isochronal} (or photometric) age from our
simulated photometry of a coeval BD sample, with typical accretion
rates and associated circumstellar discs, would be inaccurate and
exceedingly difficult. Indeed, the resulting photometric colours and
magnitudes could be indicative of a more distant, and with higher
extinction levels, CTTS population. For more extreme accretion rates
the scatter (in CMD space) is significantly removed from the naked BD
locus such that these stars have little chance of being selected as
BDs. As discussed in Section \ref{results} this does not include any
effects due to past accretion on the evolution of the central star,
which acts to accelerate the gravitational contraction and make the
star appear older \citep{tout_1999,siess_1999}, further scattering the
apparent age of a coeval population.  Concordantly, \emph{isochronal}
derivations of mass and therefore IMFs, for our simulated photometry,
of a coeval population of accreting BDs with associated discs, would
be inaccurate and problematic. Again this is caused by the changes in
the SEDs as a result of the accretion flux and increased occultation
by the larger degree of flaring seen in BD discs \citep[for the
latter, as found by][]{walker_2004}

We have also qualitatively explored the effects of accretion and disc
presence in our simulated dataset on disc fraction estimates. As is
currently well known, longer wavelength bandpasses are much more
reliable and suitable for disc identification. As shown in Section
\ref{results} the naked and BDD disc loci were much more clearly
separated in the CoCoD constructed using \textit{Spitzer} IRAC
magnitudes than the shorter wavelength CIT \textit{JHK} passbands. In
addition, we have shown that the slope of the SED, or $\alpha$ value,
from 3.6 to 8.0$\mu$m is an effective disc indicator. We have also
tentatively shown that current observational cuts, when applied to our
simulated photometry (with its associated photometric system), results
in the reliable detection of disc candidates for IRAC and MIPS colours
and $\alpha$ values, and therefore a robust lower limit disc
fraction. Cuts derived from our model grid which could be used as a
guide for observational disc candidate selection are presented in
Table \ref{cuts_table}.

A further area this study impacts on (perhaps most
significantly) is the recent evidence for a stellar mass to accretion
rate correlation, of the approximate form: $\dot{M}\propto M_*^2$
\citep{muzerolle_2003,natta_2004,natta_2006}. This relationship has
been extended into the BD mass regime in \cite{natta_2006}. However,
arguments based on selection and detection thresholds have already
cast this relation into doubt \citep{clarke_2006}. As we have shown in
Section \ref{results} a relationship of this kind is self-reinforcing
as lower mass objects with higher accretion rates have little chance
of being correctly identified as such due to both the accretion flux
and flared associated disc. Essentially, at present it is unclear how
many BDs are not included in this relationship due to
misidentification. As explained in \cite{walker_2004}, BD systems with
a disc, without including accretion effects, can have the
characteristics of higher mass CTTS stars, due to increased disc
flaring from a reduced surface gravity in the disc. The effects of
accretion at typical or larger rates further exacerbate the situation
both spectroscopically, as the photospheric flux essentially becomes
swamped or completely veiled, and photometrically as the resulting
colours and magnitudes are significantly shifted. Therefore, for our
simulated dataset a relationship of this sort may well be derived, if
typical methods are used to identify BD objects with discs and derive
masses, ages and accretion rates, even though it is not present.

Finally, although inner edge locations are correlated with their
temperature we do not find a resulting correlation with IR excess. As
our initial inner edge locations are placed at the co-rotation radius
one might expect a correlation between rotation rate and IR excess.
This in turn might suggest that studies of disc presence correlation
with slower rotation rates, exploring disc-locking, may have intrinsic
biases. However, for our systems with dust sublimation, vertical
flaring, accretion and view over a range of inclinations any
correlation is not apparent. 

We intend to extend the range of BBD models, employing both an
increased parameter space and also a set of models adopting analytical
prescriptions of the disc structure (as opposed to hydrostatic
equilibrium). We are also extending our emission line modelling
efforts \citep{kurosawa_2006} to lower central star masses in order to
identify both the location of the line emission in the light of the
work by \cite{gatti_2006} and to examine the efficacy of line fluxes
as an accretion rate probe. We are also extending our grid of CTTS
magnetospheric accretion models to lower accretion rates in order to
identify the limit at which low accretion rate CTTS are no longer
identified as accreting from H$\alpha$ line flux and morphology.

\section*{ACKNOWLEDGMENTS}
The simulations in this work were performed using the University of
Exeter supercomputer, an SGI Altix ICE 8200. NJM and TJH were
supported by STFC grant ST/F003277/1. We would also like to thank Dave
Acreman for help with the practicality of running the model grid.

\bibliographystyle{mn2e}
\bibliography{references}
\appendix

\section{Website}
\label{website}

As stated throughout this paper the data presented are available from
a web page\footnote{http://bd-server.astro.ex.ac.uk/}.

\subsection{Available Data}
\label{web_data}

The magnitudes and colours presented in this paper are available both
as individual magnitudes and as isochrones or mass tracks. Photometric
magnitudes have also been derived for several other systems and are
available online. These are Johnson, Cousins \textit{UBVRI(JHK)}
\citep{johnson_1966,bessell_2005}, \textit{Tycho} $V_t$ and $B_t$
\citep{bessell_2000}, Bessell \textit{UBVRIJHKL}
\citep{bessell_1988,bessell_1998}, SDSS \textit{UGRIZ}
\citep{fukugita_1996}, 2MASS $JHK_s$
\citep{cohen_2003,skrutskie_2006}, MKO \textit{JHK}
\citep{simons_2002,tokunaga_2002}, UKIRT \textit{ZYJHK}
\citep{hawarden_2001}, IRAS 12, 25, 60 and 100\,$\mu$m
\citep{neugebauer_1984}, SCUBA 450WB and 850WB
\citep{holland_1999} \footnote{The SCUBA 2 filter responses are similar
  to SCUBA}, Herschel PACS blue, green and red \citep{poglitsch_2008,
  poglitsch_2010} and Herschel SPIRE, 250, 350 and 500
\citep{griffin_2008,griffin_2010}. For further information on these
magnitudes, such as the filter responses used and the adopted
zeropoints please refer to the website.

In addition to the magnitudes derived for each of these bands
monochromatic fluxes have also been derived for all bands listed
above. These have been derived following closely the methods of
\cite{robitaille_2006}, extended to further passbands. For details
of the assumed SED shape, central wavelengths and bandpasses adopted
please refer to the website. The derivations of these monochromatic
fluxes will be detailed in a coming paper, which details a fitting
tool associated with these data.

\section{Consistency Checks}
\label{consistency}

Firstly, a check was made on the photospheric input flux and the
resulting stellar direct flux (tagged by {\sc torus}) after the radiative
transfer simulation. The resulting flux distributions should match
most closely for face-on configurations, and then match in shape only,
with the stellar direct flux level dropping towards higher
inclinations, as more photons are scattered and absorbed by the disc.

We also directly compared the magnitudes and colours of our naked BD
systems with the lowest accretion rate (\logmdot -12) to those
published in \cite{chabrier_2000}, in the same photometric system
(\textit{CIT}). We found significant colour differences ($\delta
(J-K)\le$ 0.1), between our derived values and those of
\cite{chabrier_2000}. As a further check we derived the magnitudes in
the \cite{bessell_1998} system (by both adopting the published zero
points, and by using a Vega reference spectrum), and applied
conversions of \cite{leggett_1992}\footnote{We have also included the
  wavelength shift mentioned in \cite{stephens_2004}}, to the
\textit{CIT} system. For each method we failed to match the magnitudes
and colours published in \cite{chabrier_2000}. As a final test we
passed the downloaded, unaltered, atmospheric spectra directly through
the filter response program, without interpolation, for the closest
matches in log(g) and $T_{\rm eff}$ from the interior models published
in \cite{chabrier_2000}. These magnitudes and colours also failed to
match. Therefore, we must conclude that the most likely cause of the
mismatch is due to improvements in the model atmospheres available
online\footnote{http://perso.ens-lyon.fr/france.allard/} (this is
likely as the models available online, have a later time stamp,
$\approx$\,2005 compared to 2000). For the final published magnitudes,
for the naked systems, we have used a similar wavelength resolution as
in our BDD systems, i.e. 200 logarithmically spaced points. This means
that magnitudes derived from these spectra will differ slightly from
those derived from the full spectra, but this effect is negligible,
and increased resolution for only some of our model grid (i.e naked
stars) will hamper comparison between the models.

The final test of the derived colours and magnitudes was a comparison
of the naked systems with the results for the almost face-on BDD
systems. The results for the optical passbands should be similar and
an appraisal of the component SED, i.e. showing the stellar direct
flux.

We have adopted zeropoints derived using a Vega reference spectrum for
the optical and near-IR passbands. As a test we have compared our
derived zeropoints using the filter response of
\cite{bessell_1998} and the Vega reference spectrum against
those published in \cite{bessell_1998}. For our photometric
systems we integrate the summed number of photons counted by the
simulated telescope systems, however to test the zeropoints and match
the system of \cite{bessell_1998} we must integrate the summed
energy. The zeropoints we derived \citep[with the values of][in
parenthesis]{bessell_1998} were: $U=$20.977(20.94),
$B=$20.499(20.498), $V=$21.116(21.10), $R=$21.676(21.655),
$I=$22.376(22.371), $J=$23.735(23.755), $H=$24.989(24.860),
$K=$25.884(26.006) and $L=$27.809(27.875). Our derived zeropoints and
those of \cite{bessell_1998} match to within 0.05 mags (and
usually much closer) for all bands except the \textit{H,K} and
\textit{L} bands. This is probably due to the previously noted IR
excess (although detected at longer wavelengths) of the observed Vega
spectrum. \cite{bessell_1998} use a combined model spectrum of
Vega and Sirius as their reference spectrum. As a further test we also
used the synthetic A0V stellar spectrum of \cite{cohen_1993}
to derive zeropoints but were still unable to improve the match to the
\cite{bessell_1998} photometric system for the \textit{JHK}
colours. However, for these colours we have adopted the \textit{CIT}
system, but were unable to find published zeropoints, and therefore
used the Vega reference spectrum. Essentially this may mean there is a
small offset in our \textit{JHK} photometry, however as most of our
results are based on differential photometry this will not affect our
conclusions.

A further complication with our adopted photometric systems is the
range of zeropoints available for the \textit{Spitzer} IRAC
photometry. For this study, as stated, we have adopted zeropoints
calculated using the zero magnitude flux from the IRAC
handbook\footnote{http://ssc.spitzer.caltech.edu/documents/som/som8.0.irac.pdf}.
The resulting zeropoints were: Channel 1[3.6]=19.541, channel
2[4.5]=19.089, channel 3[5.8]=17.395 and channel 4[8.0]=17.966. The
corresponding zeropoints derived for the MIPS passbands where: channel
1[24]=2.139, channel 2[70]=-0.2726 and channel 3[160]=-1.990.

In summary, several careful consistency checks were performed to
confirm that the resulting SEDs and photometric magnitudes behaved as
expected and matched any available published results. A failure to
match the published zeropoints in the near-IR bands of the
\cite{bessell_1998} using a Vega reference spectrum was
probably due to an IR excess in our observed Vega spectrum. However as
in general most of the conclusions or implications of this study are
based on differential photometry, this should not affect them
adversely.  Furthermore, a failure to match the published magnitudes
(and colours) for the atmospheric models in
\cite{chabrier_2000}, even using published zeropoints for the
excellently defined system of \cite{bessell_1998}, and
subsequent conversions to the required \textit{CIT} system
\citep{leggett_1992}, was prescribed to an update in the model
atmospheres available online.

\label{lastpage}
\end{document}